\documentclass[%
 reprint,
 amsmath,amssymb,
floatfix,
]{revtex4-2}

\usepackage{graphicx}
\usepackage{dcolumn}
\usepackage{bm}
\usepackage{xcolor}
\usepackage{hyperref}
\usepackage{float}
\usepackage{multirow} 

\begin{document}

\title{Optimal COVID-19 Vaccine Prioritization by Age Depends Critically on Inter-group Contacts and Vaccination Rates}

\author{Iker Atienza-Diez}
\thanks{These authors contributed equally to this work.}
\affiliation{Centro Nacional de Biotecnologia (CNB), CSIC, Madrid, Spain}%
\affiliation{%
Grupo Interdisciplinar de Sistemas Complejos (GISC)}

\author{Gabriel Rodriguez-Maroto}
\thanks{These authors contributed equally to this work.}
\affiliation{Centro Nacional de Biotecnologia (CNB), CSIC, Madrid, Spain}%
\affiliation{%
Grupo Interdisciplinar de Sistemas Complejos (GISC)}

\author{Sa\'ul Ares}
\email{saul.ares@csic.es}
\affiliation{Centro Nacional de Biotecnologia (CNB), CSIC, Madrid, Spain}%
\affiliation{%
Grupo Interdisciplinar de Sistemas Complejos (GISC)}

\author{Susanna Manrubia}
\email{smanrubia@cnb.csic.es}
\affiliation{Centro Nacional de Biotecnologia (CNB), CSIC, Madrid, Spain}%
\affiliation{%
Grupo Interdisciplinar de Sistemas Complejos (GISC)}

\date{\today}

\begin{abstract}

The limited availability of COVID-19 vaccines has prompted extensive research on optimal vaccination strategies. Previous studies have considered various non-pharmaceutical interventions, vaccine efficacy, and distribution strategies.
In this work, we address the combined effects of inter-group contacts and vaccination rates under contact reduction,
analyzing the Spanish population's demographic and age group contact patterns and incorporating reinfection dynamics.
We conduct an exhaustive analysis, evaluating 362,880 permutations of 9 age groups across 6 vaccination rates and two distinct, empirically quantified scenarios for social contacts. Our results show that at intermediate-to-high vaccination rates with unrestricted social contacts, optimal age-based vaccination strategies only slightly deviate from older-to-younger prioritization, yielding marginal reductions in deaths and infections. However, when significant reductions in social contacts are enforced ---similar to the lockdowns in 2020---, there are substantial improvements, particularly at moderate vaccination rates. These restrictions lead to a transition where infection propagation is halted, a scenario that became achievable during the pandemic with the observed vaccination rates.
Our findings emphasize the importance of combining appropriate social contact reductions with vaccination to optimize age-based vaccination strategies, underscoring the complex, nonlinear dynamics involved in pandemic dynamics and the necessity for tailored, context-specific interventions.
\end{abstract}

\maketitle

\section{Introduction}

Compartmental models have been broadly used to characterize the dynamics of infection propagation in a variety of scenarios. Though most studies have addressed the impact of refining basic susceptible-infected-recovered dynamics \cite{kermack:1927} to include a multiplicity of additional states \cite{hethcote:2000}, others have thoroughly explored the effect of various external interventions.
The use of such models has become widespread, addressing issues such as the effect of contention measures \cite{arenas:2020,wong:2020-PRX,ventura:2022}, of the immune state of the population \cite{weitz:2020}, of population density \cite{wong:2020}, or of vaccine rollout \cite{wagner:2022}, among others. One of the major strengths of modeling approaches is that they offer the possibility of comparing various scenarios under similar conditions. Altogether, such studies have refined our understanding of epidemic dynamics and of the role played by multiple relevant variables. 

The availability of a vaccine against SARS-CoV-2 largely before the infected population attained a sufficiently high immunity level to halt propagation prompted the study of optimal protocols to administer available vaccine doses \cite{matrajt:2021,schulenburg:2022,gonzalez-parra:2024}. Due to the high, positive correlation of SARS-CoV-2 mortality with age of the infected individual, there was a largely preferred protocol of administration where vaccination of the elderly was prioritized, to continue in decreasing age order once most vulnerable groups had been protected. Results obtained through models incorporating a variety of scenarios, but implementing COVID-19 features, systematically supported that mortality and years of life lost were minimized if the population over 60 to 70 years old was given priority \cite{bubar:2021,tran:2021,zhao:2021,gonzalez-parra:2024,nguyen:2023}.
Some of these studies used data from sociological studies elaborated in the absence of social contact limitations or, in the best case, implemented such limitations
%
%
in a phenomenological way, using for instance effective values of the reproduction number in the simulations \cite{bubar:2021,zhao:2021}.
However, modifications in social contact patterns during an epidemic event can change due to both non-pharmaceutical interventions and as a result of unsupervised responses of individuals~\cite{pan:2020,manrubia:2022} in a highly heterogeneous way \cite{caselli:2022}. During the COVID-19 pandemic, patterns of contact between individuals were severely modified \cite{ayouni:2021}, differently affecting citizens involved in public services, children, or elders, for example. 

In this contribution, we carry out an exhaustive exploration of age-based vaccination protocols by measuring the reduction in deaths and infections obtained under various scenarios. Our baseline situation considers an age-structured population with 9 age groups and a specific demographic structure: the Spanish population. In our first scenario, no restrictions to social contacts are implemented, though empirical data regarding inter- and intra-group contacts is used \cite{mistry:2021}. Under the previous conditions, we carry out simulations with a compartmental model that takes into account partial immunity and, therefore, reinfection of individuals, as well as vaccination at various rates.
All possible orderings of the population, attending to the age of the different groups defined in our study, are simulated.
This is a total of $9!=362,880$ possibilities for each vaccination rate. As a result, we are able to identify the best possible protocol and quantify its benefits with respect to the baseline case. The whole study is repeated when lockdown measures are implemented, using the results of two studies \cite{backer:2021,gimma:2022} that have measured contact reduction in different age groups. Our findings align with results in the literature under no contact restrictions, showing clearly that the largest benefit in terms of mortality reduction is obtained when vaccination prioritizes elderly people. However, we obtain a substantial advantage of vaccinating first people in their thirties under lockdown conditions. In this second situation, reductions in mortality and in number of infections can be simultaneously optimized under similar vaccinating strategies. We also show that there is a non-trivial interaction between the degree of social contacts and the vaccination rate, and that only a suitable combination of both variables can yield the largest beneficial effects. 

\section{Model and Data}

\subsection{SIYRD: a model with reinfection and vaccination}

In a previous study \cite{rodriguez-maroto:2023}, we introduced SIYRD, a compartmental model with five different classes: Susceptible (S), Infected (I), Reinfected (Y), Recovered (R) and Dead (D) individuals. The model implemented two main novelties: vaccination (S individuals move to class R upon vaccination) and reinfection (R individuals can move to the reinfected class Y). Only vaccination of susceptible individuals is considered, and an equivalence between vaccination and disease overcoming regarding the immune state is assumed. Empirical studies have found no significant differences in recovered individuals between the immune effects elicited by vaccination or by infection \cite{ons:2021,shenai:2021}, thus supporting this hypothesis.

Numerical and analytical studies of the SIYRD model showed that the introduction of reinfections causes the appearance of very long transients where the disease behaves as quasi-endemic.
Extending the model to include a synthetic population divided into two age groups revealed that at high vaccination rates ---whenever disease impact intensifies with age--- the optimal vaccination strategy shifts from prioritizing the elderly to prioritizing younger individuals.
Here, our main goal is the exhaustive exploration and quantification of this observed effect in realistic scenarios.

To this end, we extend the SIYRD model to encompass 9 different age groups. This allows us to implement the demographic structure of the population and inter- and intra-group empirical contact matrices in two situations: without and with limitations to social contacts. The numerical analysis uses data from Spanish demography and COVID-19 data, though it is easily applicable to other diseases, countries or group-contact scenarios. 

\subsection{S$^9$IYRD: a model implementing demographic and group-contact structure}

Let us define $S_i$ as the number of susceptible individuals in age group $i$, with $i=1$ including from newborns to individuals of age 9, $i=2$ individuals between ages 10 and 19, and so on until $i=8$. The last group, $i=9$ includes all individuals of age 80 and older. The number of individuals for other classes and all age groups are analogously described by $I_i$, $Y_i$, $R_i$ and $D_i$, $i=1, \dots 9$.

\begin{figure}[t]
\begin{center}
 \includegraphics[width=0.45\textwidth]{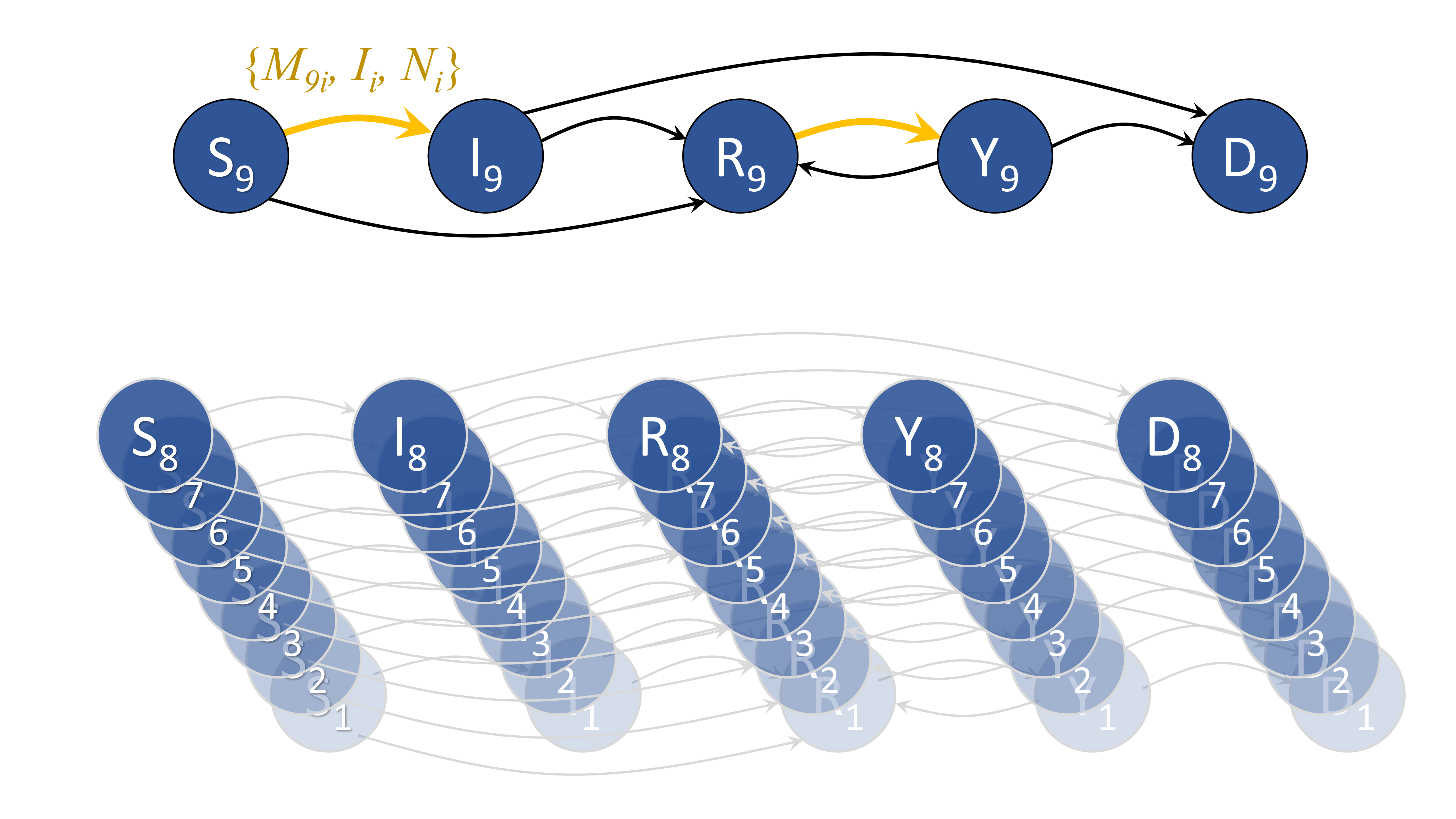}
\caption{Schematic of the S$^9$IYRD model. Five different compartmental classes are considered: susceptible, S; first infected, I; recovered (either recovered from I or vaccinated from class S), R; reinfected (infected after vaccination or after recovering from a first infection), Y; dead, D. Individuals are further classified in one out of nine different age groups. Contacts between infected individuals (I or Y) and those that can be infected (S or R) occur between all possible age pairs $\{i,j\}$ with a weight depending on contagion parameters, demographic structure and empirical contact matrices (orange arrows in the schematic). 
}
\label{fig:S9IYRD}
\end{center}
\end{figure}   

Equations (\ref{eq:siyrd-Si}-\ref{eq:siyrd-Di}) describe the dynamics of each of the nine groups in the S$^9$IYRD model, as schematically shown in Fig.~\ref{fig:S9IYRD},

\begin{align}
    \label{eq:siyrd-Si}
        \dot{S_i} &= - \left[ \sum_{j=1}^{9}\left( \beta_{SI}I_j +\beta_{SY} Y_j \right) \frac{M_{ij}}{N_j} \right] S_i - v_i \Theta (S_i,\theta) \\
    \label{eq:siyrd-Ii}
        \dot{I_i} &= \left[ \sum_{j=1}^{9} \left( \beta_{SI} I_j + \beta_{SY} Y_j \right) \frac{M_{ij}}{N_j} \right] S_i - (r_i + \mu_{I_i}) I_i \\
    \label{eq:siyrd-Yi}
        \dot{Y_i} &= \left[ \sum_{j=1}^{9} \left( \beta_{RI} I_j + \beta_{RY} Y_j \right) \frac{M_{ij}}{N_j} \right] R_i - (r_i + \mu_{Y_i}) Y_i \\
    \label{eq:siyrd-Ri}
            \dot{R_i} &= \begin{aligned}[t]
            &- \left[ \sum_{j=1}^{9} \left( \beta_{RI} I_j + \beta_{RY} Y_j \right) \frac{M_{ij}}{N_j} \right] R_i \\
            &+ r_i (I_i + Y_i) + v_i \Theta (S_i,\theta)
        \end{aligned} \\
    \label{eq:siyrd-Di}
        \dot{D_i} &= \mu_{I_i} I_i + \mu_{Y_i}Y_i
\end{align}

Parameters are rescaled such that the time unit of numerical simulations is one day. The meaning of parameters and the estimated values based on empirical data are discussed in the next subsection. 

\subsection{S$^9$IYRD model parameters}

Parameter values chosen for our numerical simulations consider the early propagation of COVID-19 and the Spanish population as an example. A summary of the main characteristics of the latter used to feed model parameters is represented in Fig.~\ref{fig:Spain_9G}. Though the qualitative results obtained are robust to variations in virus and population characteristics, quantitative results would need to be reevaluated for different situations. Omicron, for instance, is characterized by a milder impact fatality rate (thus yielding different mortality and recovery rates) \cite{adjei:2022,wang:2023}, a different infectious period, and higher infection and reinfection rates than early COVID-19 circulating strains. Also, epidemic dynamics should differ between countries with different social contact habits, or regions with an expansive population pyramid ---unlike the constrictive Spanish population pyramid. 

\subsubsection{Infection rates}

The separation between primary infections (I) and reinfections (Y) entails four different infection rates: $\beta_{SI}$ and $\beta_{SY}$ correspond to the infection rates of susceptible individuals due to contacts with primary-infected (I) and re-infected (Y) individuals, respectively; $\beta_{RI}$ and $\beta_{RY}$ are the analogous rates for recovered individuals.
Note that since we do not contemplate an Exposed class in the model, incubation periods are effectively included in the infection and recovery rates.
Although the precise values of some of these infection rates might be difficult to estimate in general, there are sensible relationships among the parameters that should hold, on average, over the population. We assume individuals in the reinfected class Y bear a lower viral load due to their previous exposure to the virus, and therefore are less infective than primary-infected individuals in class I, both towards susceptible S and towards recovered R individuals. This implies that reinfection rates are smaller than primary-infection rates, $\beta_{SI} \geq \beta_{SY}$ and $\beta_{RI} \geq \beta_{RY}$. 

In the same vein, the likelihood that a susceptible individual becomes infected is larger than that of a recovered individual, since the latter bears at least partial immunity against the disease either due to prior infection or to vaccination. This applies both to primary infections, $\beta_{SI} \geq \beta_{RI}$, and to reinfections, $\beta_{SY} \geq \beta_{RY}$. Finally, we assume that the ratio between the infection rates of primary-infected and reinfected individuals is independent of the state of the individual that can be potentially infected,

\begin{gather}
\label{eq:par_rel}
\frac{\beta_{SY}}{\beta_{SI}}=\frac{\beta_{RY}}{\beta_{RI}} \, .
\end{gather}

Lacking specific data that suggest otherwise, we assume that COVID-19 transmission rates $\beta_{SI}$, $\beta_{SY}$, $\beta_{RI}$ and $\beta_{RY}$ are independent of the age group considered. Precise values can depend on different COVID-19 strains, but are largely country-independent. 

Specifically, infection rates were estimated as:
\begin{align}
\label{eq:betas}
\beta_{SI} & =\frac{R_{SI}}{d_I} \qquad \beta_{RI}=\alpha_1\beta_{SI} \qquad \nonumber \\
\beta_{RY} & =\alpha_2\beta_{RI} \qquad \beta_{SY}=\frac{\beta_{RY}\cdot\beta_{SI}}{\beta_{RI}} \, 
\end{align}
where $R_{SI}$ parametrizes the transmissibility of the epidemic disease under consideration, $d_I$ is the infectious period (the time interval during which the individual is infectious), and $\alpha_i$ are two constants satisfying $\alpha_i \leq 1$. 

In general, the reproductive number $R(t)$ varies along epidemic propagation, usually starting at high values at the beginning of an epidemic wave and decaying toward one as the epidemic progresses. Such has been the case with the COVID-19 reproductive number, which has been observed to fluctuate all over the world around its critical transmissibility value $R=1$, with occasional departures at the emergence of new strains~\cite{manathunga:2023} or when social contact
restrictions are put in place~\cite{koyama:2021}.
Under free epidemic propagation, $R(t)$ has consistently tended towards 1 \cite{arroyo:2021}, probably due to changes in individual risk aversion in response to the epidemic state~\cite{manrubia:2022}. In view of this evidence, for simplicity we choose to fix $R_{SI} = 1$, as representing the baseline long term trend in $R(t)$.

The infectious period $d_I$ results from the sum of two contributions: the average exposure period of infected individuals, estimated at $3$ days \cite{bubar:2021}, and the infectious period proper. Data suggest that patients with mild to moderate early COVID-19 remain infectious no longer than 10 days after symptom onset \cite{website:d_I}. Therefore, we take $d_I = 13$ days. 

Finally, we have chosen $\alpha_1=0.01$, modeling the first COVID-19 wave \cite{murchu:2022} before significantly mutated strains appeared, and $\alpha_2=0.5$, implying that reinfected individuals recover twice as fast as primary-infected individuals. 

\subsubsection{Mortality and recovery rates}

\begin{figure}[t]
    \centering
    \includegraphics[width=0.45\textwidth]{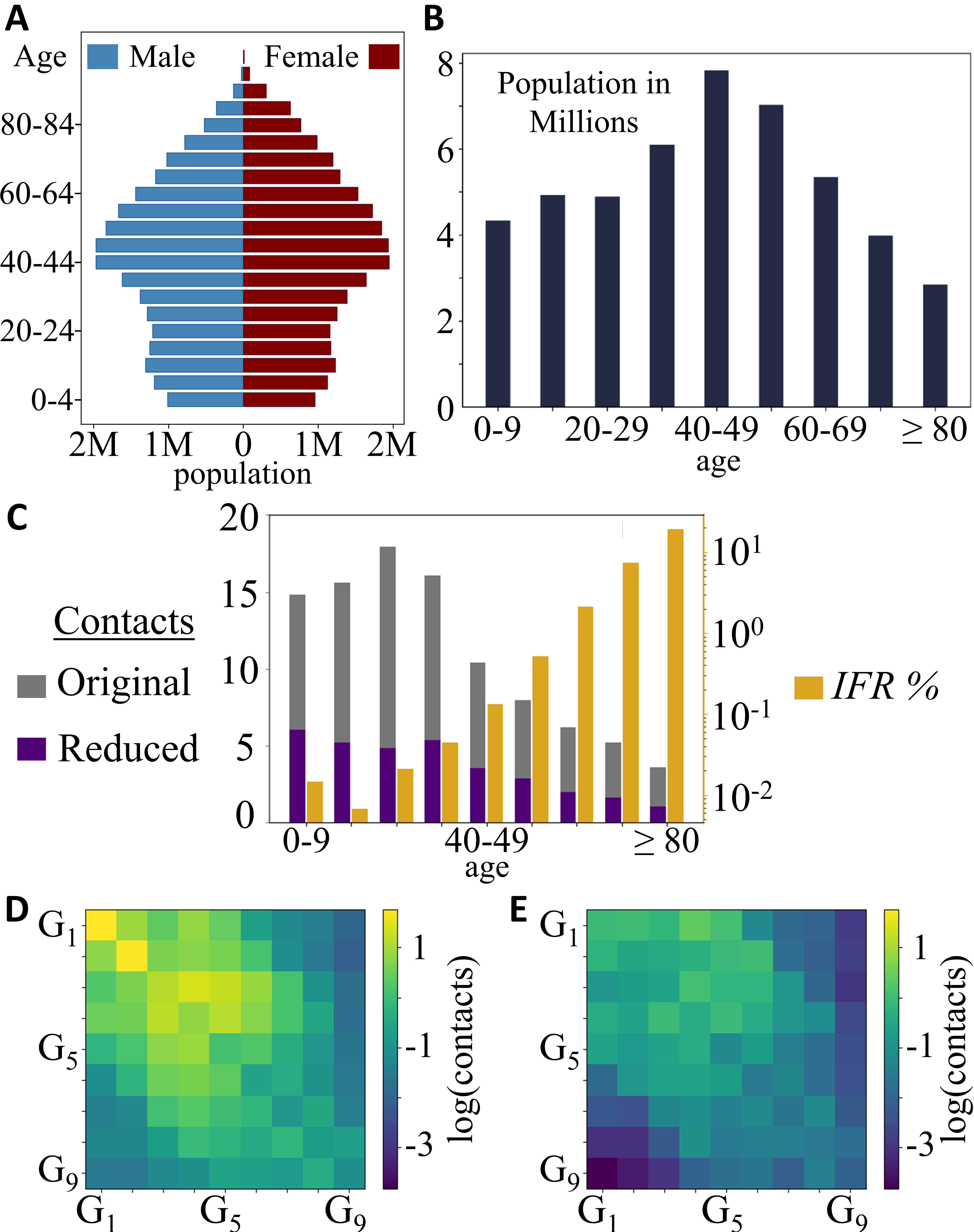}
    \caption{S$^9$IYRD parameters for COVID-19 in Spain. (A) Demographic pyramid for the Spanish population in 2020; blue corresponds to male population, dark red to female population; in the vertical axis, populations are grouped in 5-year intervals, starting with 0-4 years at the bottom. (B) Population size considering nine age groups in 10-year intervals. (C) Effective values of the number of daily contacts for each group, considering the original (gray) and reduced (purple) contact matrix. In gold, (log scale, axis on the right side) we represent the infection fatality rate for each age group in Spain \cite{website:isciii}). (D) Heatmap of the contact matrix representing the logarithm of the average number of contacts between individuals \cite{mistry:2021} for the nine groups-stratified Spanish population. (E) Reduced contact matrix, as in (D) but in a scenario with limitations to the number of social contacts (see text for details).}
    \label{fig:Spain_9G}
\end{figure}

The quantitative estimation of the primary mortality rate depends on the infection fatality rate (IFR) and on the infectious period of a disease. The values of the IFR for COVID-19 in Spain \cite{website:isciii} are represented in Fig.~\ref{fig:Spain_9G}C, though measures in different world countries return comparable values~\cite{spiegelhalter:2020,odriscoll:2021}. For primary-infected individuals, $\mu_{I_i}$ can be calculated from the IFR in each age group, $IFR_i$, and the infectious period $d_I$ as

\begin{align}
\label{eq:mu_I_i_IFR}
    \mu_{I_i} &=  \frac{IFR_i}{d_I} \, ,
\end{align}
where $IFR_i$ is given by the ratio between the number of fatalities and the number of infections in a given group $i$. The recovery rate from a primary infection is analogously estimated from

\begin{align}
    r_i &=  \frac{1-IFR_i}{d_I} \, .
\label{eq:r_i_IFR}
\end{align}

Mortality rates of secondary infections, $\mu_{Y_i}$, are set to $0$ for all nine groups. This, by definition, establishes an endemic disease where, asymptotically, all individuals that have either been vaccinated or have survived a primary infection belong to R or to Y classes. Considering that this model does not incorporate the possibility of mutations in the circulating strain, assuming that secondary infections do not cause fatalities seems sensible. Still, this limits the applicability of the model to long time intervals, where changes in the health state of individuals or in the circulating virus might take place. Since the goal of the model is to establish optimal vaccination protocols in a mostly susceptible population (at the onset of the disease propagation), a  timescale of about one year suffices, thus justifying the study of the limit case $\mu_{Y_i}=0$. Finally, we assume that the recovery rate $r_i$ of primary-infected and reinfected individuals is identical, neglecting possible small differences \cite{hadley:2023,nguyen:2023}. 

\subsubsection{Vaccination rates} 

Vaccination is implemented through a parameter $v_i$ that represents the fraction of susceptible individuals in group $i$ vaccinated per time unit. 
The rate is multiplied by a function $\Theta(S_i,\theta_i)$ to indicate that vaccination is achieved only in a fraction $1-\theta_i$ of individuals in that group. Since it has been shown that qualitative results are independent of the specific functional form of $\Theta(S_i,\theta_i)$ \cite{rodriguez-maroto:2023}, we assume that vaccination takes place at a constant rate until the fraction $1-\theta_i$, with $\theta_i = \theta=0.3$ for all groups $i$, is achieved. 

We explore priority vaccination protocols determined by a specific order of the nine age groups. Specifically, the prioritized population group is vaccinated at a rate $v$ until the fraction of susceptible individuals reaches $(1-\theta)$ in that group. When that happens, vaccination of the next prioritized group begins ---provided the previous threshold $\theta$ has not yet been reached through natural infections, a situation that may hold for too low values of $v$---. Vaccination continues with other groups until the vaccination protocol is completed.
The vaccination rate $v_i=v$ is a variable that can be explored in a range of values depending on vaccine supply. We consider 6 different vaccination rates: $0.05$, $0.1$, $0.5$, $1.0$, $1.5$ and $2\%$ representing the fraction of population vaccinated per day. 

In the case of the Spanish population, with a total around $47M$ individuals, slow vaccination rates of $0.05$ and $0.1$ would correspond to $23,500$ and $47,000$ daily vaccines administered, while the fastest rate considered in the simulations, $v=2.0$, implies that $940,000$ individuals would be vaccinated per day. The highest 7-day average daily vaccine administration actually achieved ---during the Omicron wave in late $2021$--- was $1.41\%$ \cite{website:vaccESP}.

\subsubsection{Demographic structure and social contacts}
\label{sec:constrainedcontacts}

The demographic structure of the population, Fig.~\ref{fig:Spain_9G}A, sets the size $N_i$ of each age group, Fig.~\ref{fig:Spain_9G}B, a quantity that serves to normalize the value of $M_{ij}$. 
The elements $M_{ij}$ of the contact matrix $\bf M$ are the number of contacts an individual of group $i$ has with individuals of group $j$, thus weighing the effect of intra- ($j=i$) and inter- ($j \ne i$) group contacts in contagion rates. 

We have used different sources to quantify contact matrices in a situation where contacts are not limited and when there are contact restrictions. In the former case, with no restrictions to contacts,
data for Spain could be obtained and used as found in a previously published comprehensive study~\cite{mistry:2021}, see also Fig.~\ref{fig:Spain_9G}D. Data on contact reduction under lockdown for Spain were not available, however, so we resorted to two independent studies for the Netherlands~\cite{backer:2021} and England~\cite{gimma:2022} to estimate a realistic matrix under contact reduction. Both studies compared the reduced matrix to a baseline, pre-pandemic situation. By using the difference between contacts under lockdown versus contacts without restrictions, we illustrate how lockdown measures impact depends on the age group, affecting in turn optimal vaccination protocols. 

The study with the Dutch population~\cite{backer:2021} is a cross-sectional survey in which participants reported the number and age of their contacts the previous day. It was conducted at three different times; in this work, we are interested in two of these times, namely, data collected from Feb 2016 to Oct 2017 (pre-pandemic) and data collected in Apr 2020, after strict physical distancing measures were implemented. These included closing daycare centers, schools, universities, caf\'es, pubs, restaurants, theaters, cinemas and sport clubs, as well as canceling events with more than 10 participants. The advice to citizens was to work from home whenever possible and to maintain 1,5m distance from others outside the household. Overall, the effective number of contacts per person was found to be reduced by 69\%. The distancing measures taken are very similar, only slightly more permissive, than those enforced in Spain in Apr 2020. Data in~\cite{backer:2021} can be directly applied to our case once we have taken care of the differences in age grouping, which simply implies a recalculation of the number of contacts per age group using weights given by the size of each group. 

The baseline for the effect of lockdown measures in the English study~\cite{gimma:2022} was a previous population-based prospective survey of mixing patterns carried out in eight European countries~\cite{mossong:2008}. Interestingly, this study showed that mixing patterns and contact characteristics were remarkably similar across countries (Spain was not included in the study). From March to June 2020, lockdown in England entailed work from home, schools closed, restaurants closed and mandatory masks in some areas. The mean number of contacts was about 75\% less than at pre-pandemic time~\cite{gimma:2022}. Data had to be processed again just to meet the age grouping used in our study. 

Reduced matrices obtained from the Dutch and the English study show qualitatively similar patterns and are quantitatively comparable. They can be found in the file SuppFile\_ConstrainedMatrix.xlsx, available at \cite{Atienza2024CodeZenodo}. We have averaged the two values obtained for each of the ${\bf M}$ matrix elements, resulting in values that we have used in our simulations as a case-example of contacts under restrictions (see Appendix \ref{appendix:M}). The final matrix that we implemented is represented in Fig.~\ref{fig:Spain_9G}E. 

\subsection{S$^9$IYRD numerical implementation}

To account for the possibility of different emerging timescales in the dynamics due to action of various mechanisms in the model, we have numerically integrated the equations using the variable-step, variable-order (VSVO) implicit solver ode15s from the MATLAB ODE suite \cite{shampine:1997}. This solver is appropriate for stiff problems with different timescales. We have not observed unrealistic crossings of variables to negative values \cite{satsuma:2004}. It also performs well for non-stiff problems, so it is a safe choice for compartmental models that tend to be well-behaved regarding numerical simulations.
Data and relevant code for this research work are stored in GitHub \cite{Atienza2024Code} and have been archived within the Zenodo repository \cite{Atienza2024CodeZenodo}.

Given the fixed parameters (all rates specified in previous sections), a vaccination rate and a social scenario (unrestricted or reduced physical contact), simulations are run for a time interval corresponding to one year using all possible permutations for vaccination of age groups. Each vaccination protocol (permutation of the 9 age groups) is written as a vector with a specific group order. For example, $[7, 8, 9, 1, 2, 3, 6, 5, 4]$ indicates that group 7 (individuals from 60 and 69 years old) is vaccinated in the first place, followed by group 8, then 9, and so on until either the simulation finishes or 70\% (or over) of individuals in a given group have been infected. This latter case is common at low vaccination rates; for example, with the parameters chosen and for $v=0.05$, and if vaccination starts with the oldest group, all other groups reach a fraction of over 70\% infected individuals before their vaccination turn arrives. We represent this situation as $[9, 0, 0, 0, 0, 0, 0, 0, 0]$ to indicate that only group 9 could be vaccinated, while the others reached their ``immunity thresholds'' due to natural infections. There are $9!=362 \, 880$ different orderings, which are exhaustively explored in this contribution. The efficacy of each ordering (each vaccination protocol) is quantified through the percent reduction RD\% in the total number of deaths and the percent reduction RI\% in the total number of infections, as compared to the baseline of no vaccination. 

\begin{figure}[t]
    \centering
    \includegraphics[width=0.45\textwidth]{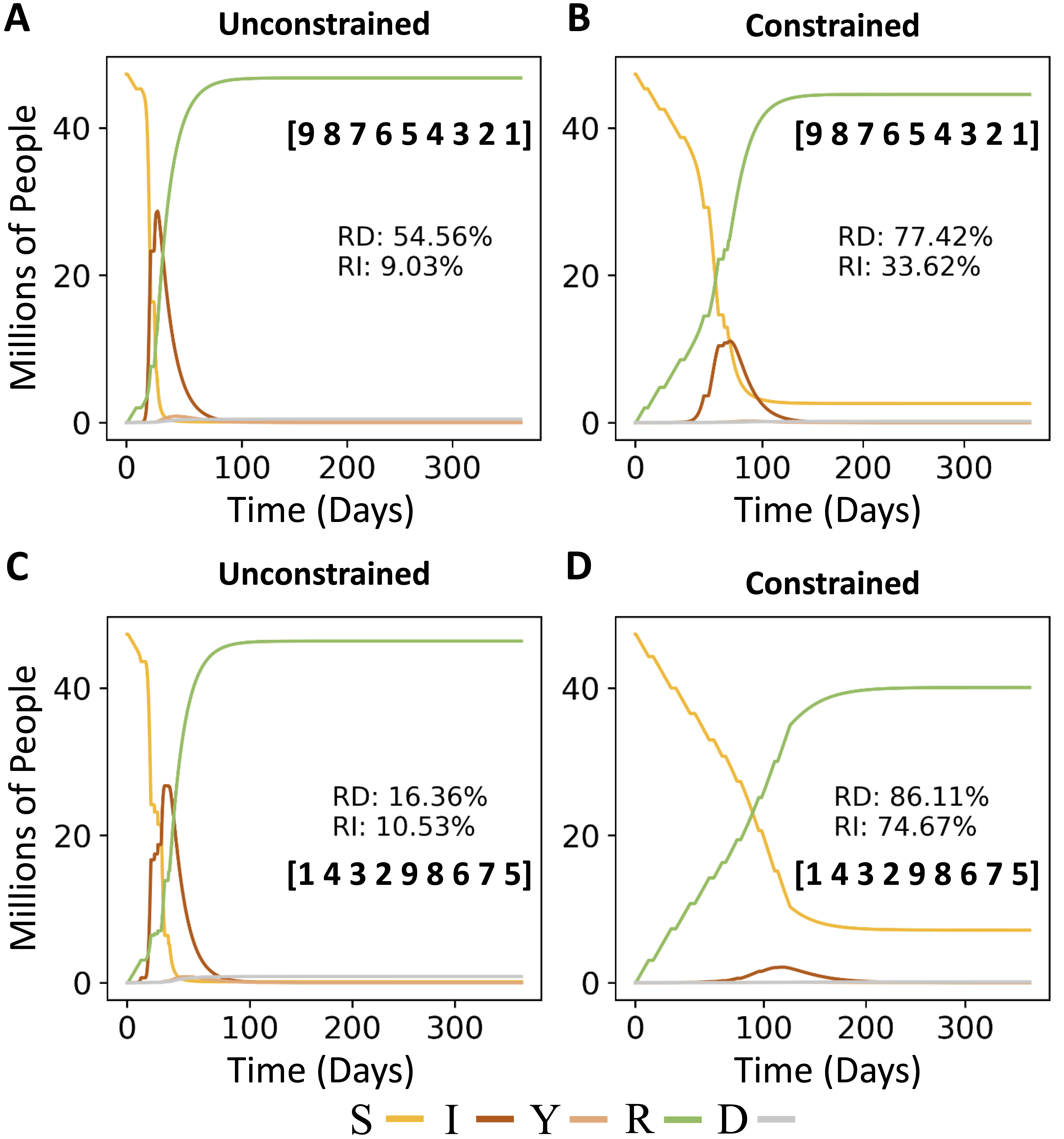}
    \caption{Dynamics of the S$^9$IYRD model in representative scenarios. The $x$-axes represent time in days since the start of epidemic propagation. The inset values show the percentage reduction in deaths (RD) and infections (RI) at the simulation endpoint (one year), calculated relative to the baseline scenario of no vaccination. In this figure, the vaccination rate is $v = 0.5$. The two upper plots illustrate the dynamics under an older-to-younger vaccine administration protocol, while the two lower plots correspond to dynamics using a representative protocol that maximizes RD at intermediate vaccination rates 
    (see Table \ref{table:redMx_bestRD_RI}).
    The scenarios depicted are: (A) unconstrained social contacts with an older-to-younger vaccination protocol; (B) constrained social contacts with the same vaccination protocol as in (A); (C) unconstrained social contacts with the protocol $[1,4,3,2,9,8,6,7,5]$; and (D) constrained social contacts with the same vaccination protocol as in (C).}
    \label{fig:ExampleDynamics}
\end{figure}

Figure \ref{fig:ExampleDynamics} shows the dynamics of the S$^9$IYRD model in four representative scenarios, using the parameters described earlier. This figure demonstrates how a well-coordinated combination of social contact reduction and an optimal age-based vaccination protocol can significantly reduce the total number of fatalities and infections (compare Fig. \ref{fig:ExampleDynamics}A and \ref{fig:ExampleDynamics}D). Additionally, the number of simultaneously infected individuals is much lower in the latter scenario, thereby lessening the strain on the healthcare system. These dynamics will be systematically explored in the following section.

\section{Results}

\subsection{Optimal ordering under unconstrained contacts}

We have first explored the situation of epidemic propagation using a contact matrix that reflects the behavior of a population when no limitations to social contacts are imposed~\cite{mistry:2021}. This scenario has been explored in various studies~\cite{bubar:2021,tran:2021}, despite the fact that assuming no changes in contact behavior is unrealistic when a threatening contagious disease is spreading: even in the absence of imposed institutional measures, individual behavioral changes typically arise as a result of the perceived risk of contagion~\cite{ferguson:2007,funk:2010,butler:2014}. Still, the case of no changes in measures that hinder contagion, spontaneous or imposed, serves as a baseline to compare different scenarios under otherwise identical conditions. 

\subsubsection{Reduction in fatalities and infections} 

\begin{figure}[t]
    \centering
    \includegraphics[width=0.4\textwidth]{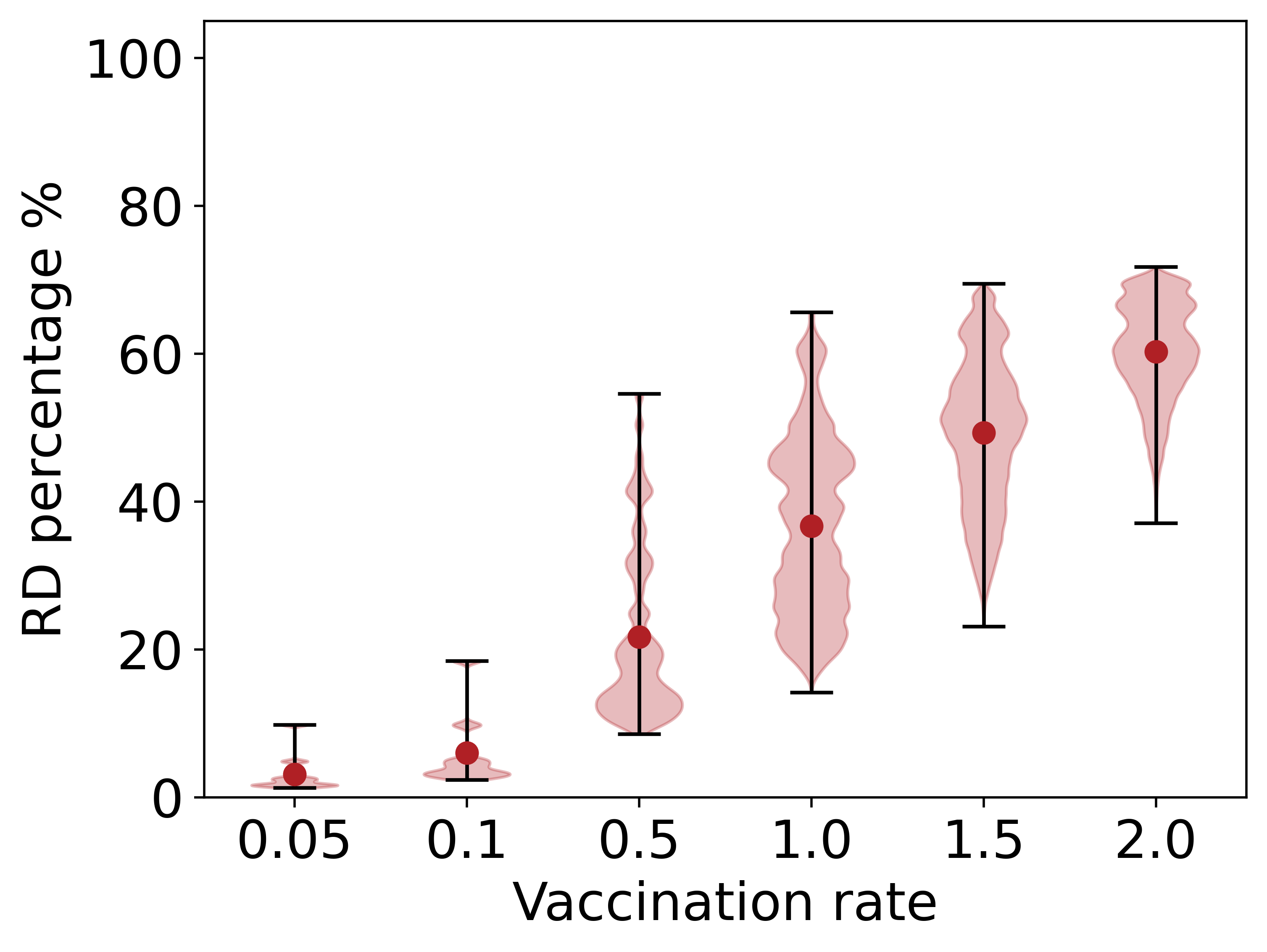}
    \includegraphics[width=0.4\textwidth]{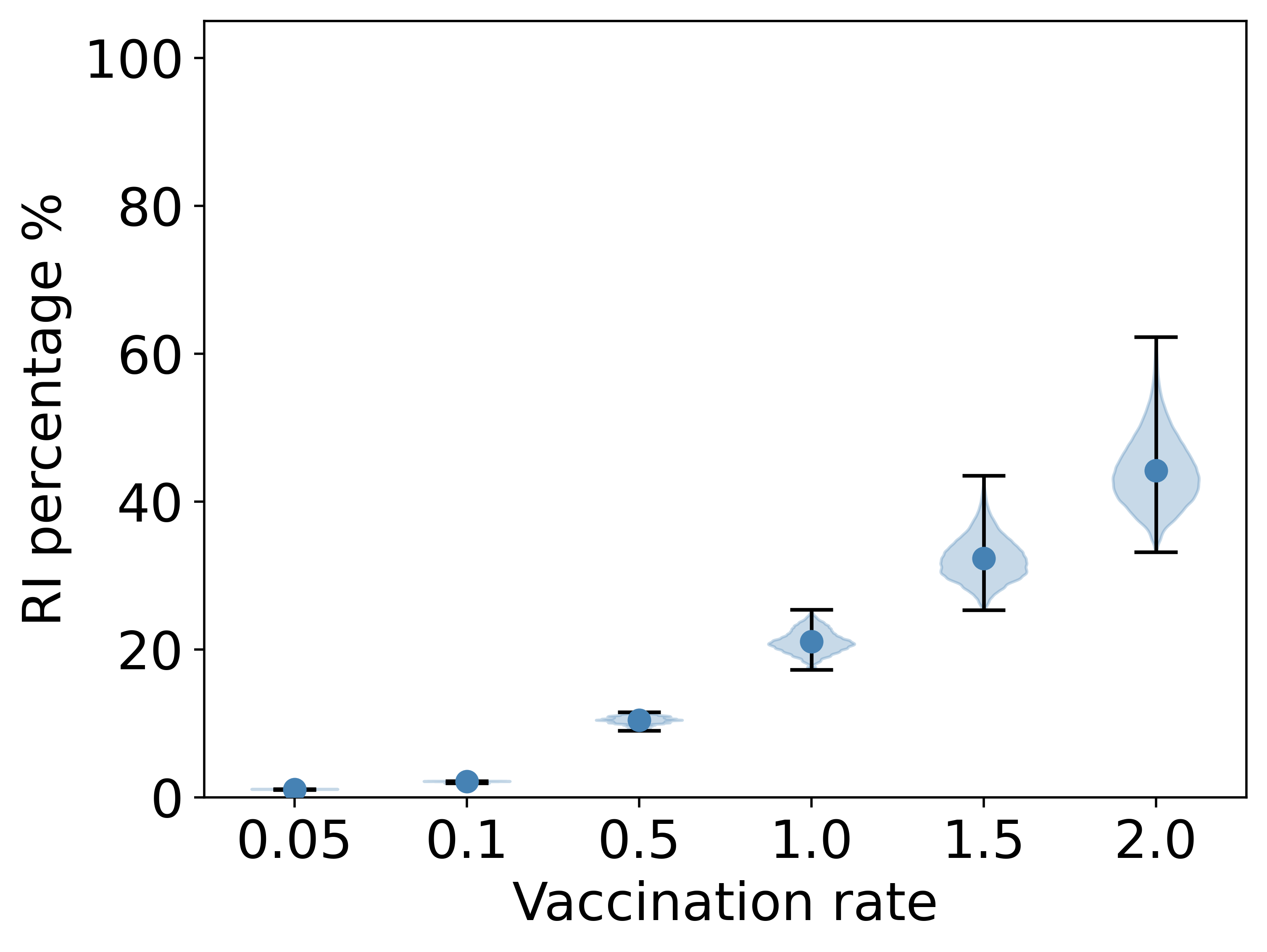}
    \caption{Distribution of the percentage reduction in deaths (top) and infections (bottom) for all possible permutations of the vaccination protocol, for each vaccination rate, in the case with unconstrained contacts. Dots represent average values, and black bars span from maximum to minimum effect.}
    \label{fig:violinUnconstrained}
\end{figure}

The percent reduction in the number of fatalities and infected individuals depends significantly on the vaccination rate and on the vaccination ordering of age groups. We present a statistical summary of these reductions in Fig.~\ref{fig:violinUnconstrained} through a violin plot of RD\% and RI\% values, for the ensemble of all possible orderings. As it could have been expected, increasing the vaccination rate improves, on average, the effect of vaccination, diminishing both the total number of deaths and of infections. It is important to note that a bad choice of the vaccination protocol can however suppress the advantage of increasing the vaccination rate, affecting especially the total number of deaths.

Too low vaccination rates, $v \le 0.1$ change only slightly the course of the epidemic, which is dominated by its unconstrained propagation. However, the effect of vaccination changes drastically for larger rates: for $v=0.5$ and higher, the order of the administration protocol can cause benefits that vary from about only 10\% improvement (both in deaths and infections) to over 50\% decrease in total number of fatalities when the optimal ordering is chosen. 

\subsubsection{Optimal protocols differ from elder-to-younger ordering}

While for an insufficiently high vaccination rate the specific ordering of vaccination of age groups does not significantly affect the total number of deaths and infections, the situation changes as $v$ increases. For average vaccination rates, $v=0.5$ and $v=1$, there is an optimal ordering to minimize the number of deaths that coincides with the strict age ordering from older to younger groups $[9,8,7,6,5,4,3,2,1]$, widely favored for COVID-19. However, for the higher rates tested, $v=1.5$ and $v=2$, there are protocols that outperform the oldest-to-youngest (SAO strategy) ordering.
Optimal strategies imply an important reduction in infections with respect to  strict age ordering, although the results are similar in terms of RD\%, see Table~\ref{table:optimalUnconstrained}.
At any vaccination rate, we observe that the protocol that minimizes the number of fatalities does not coincide with the protocol that minimizes the number of infections, as it can be seen by comparing the upper (RD) and lower (RI) half of Table~\ref{table:optimalUnconstrained}.  

\begin{table}[t]
\centering
    \caption{Optimal vaccination orderings to minimize the number of deaths and the number of infections under no contact restrictions. The upper part of the table (Best RD) shows, for different vaccination rates $v$ (second column), the protocol (third column) that maximizes RD\%, the corresponding RD\% (fourth column) and RI\% (fifth column);
    for sake of comparison, the last two columns show RD\% and RI\% for the same vaccination rates and the strict age order (SAO) protocol $[9,8,7,6,5,4,3,2,1]$.
    The lower part of the table (Best RI) is analogous, but showing now the protocols that maximize RI\%.}
    \label{table:optimalUnconstrained}
    \begin{tabular}{|c|l|c|r|r|r|r|}
        \hline
        {\bf } & $v$ & Protocol & RD\% & RI\% & SAO RD\% & SAO RI\% \\
        \hline 
        \multirow{6}{0.75em}{\rotatebox[origin=c]{90}{Best RD}}
        & $0.05$ & [9 0 0 0 0 0 0 0 0] & $9.8$ & $1.0$ & $9.8$ & $1.0$ \\
        & $0.1$ & [9 8 0 0 0 0 0 0 0] & $18.5$ & $1.9$ & $18.5$ & $1.9$ \\ 
        & $0.5$ & [9 8 7 0 0 0 0 0 0] & $54.6$ & $9.0$ & $54.6$ & $9.0$ \\
        & $1.0$ & [9 8 7 6 0 0 0 0 0] & $65.6$ & $17.3$ & $65.6$ & $17.3$ \\ 
        & $1.5$ & [7 9 8 6 5 0 0 0 0] & $69.4$ & $25.5$ & $69.2$ & $25.3$ \\
        & $2.0$ & [3 2 4 9 8 7 6 5 0] & $71.8$ & $51.2$ & $70.7$ & $33.2$ \\ \hline \hline
        \multirow{6}{0.75em}{\rotatebox[origin=c]{90}{Best RI}} 
        & $0.05$ & [3 2 1 5 6 7 8 9 0] & $1.3$ & $1.1$ & $9.8$ & $1.0$ \\ 
        & $0.1$ & [3 2 4 6 7 8 9 0 0] & $2.5$ & $2.2$ & $18.5$ & $1.9$ \\ 
        & $0.5$ & [3 4 5 6 7 8 9 0 0] & $8.7$ & $11.5$ & $54.6$ & $9.0$ \\
        & $1.0$ & [3 4 2 1 5 6 7 8 9] & $14.6$ & $25.4$ & $65.6$ & $17.3$ \\ 
        & $1.5$ & [3 2 4 1 5 6 8 7 9] & $28.0$ & $43.5$ & $69.2$ & $25.3$ \\
        & $2.0$ & [3 2 4 1 5 6 7 8 9] & $52.9$ & $62.3$ & $70.7$ & $33.2$ \\ 
        \hline
    \end{tabular}   
\end{table}

\subsubsection{Many age-ordering protocols yield similar advantages}

As the spread of RD\% and RI\% values in Fig.~\ref{fig:violinUnconstrained} shows, there are significant quantitative differences among protocols. In this section, we explore in further detail how similar are orderings that perform comparably regarding the reduction caused in the two former variables. We focus our analysis in the subset of protocols that cause a reduction of 95\% or higher than the optimal protocol, both in RD\% and RI\%. We recall here that, for each vaccination rate, we have analyzed $9! = 362 \, 880$ different orderings; the top 5\% performance includes several thousand different possibilities. It is important to clarify that not all these possibilities are different in practice. For low vaccination rates, in particular, epidemic propagation proceeds faster than vaccination, so that just one or two groups can be vaccinated (before all the groups reach the 70\% threshold). Therefore, what we call protocol $[9,0,0,0,0,0,0,0,0]$ (as in Table~\ref{table:optimalUnconstrained}, best RD for $v=0.05$), actually contains $8!=40 \, 320$ protocols (corresponding to all possible orderings of groups 8 to 1 from the second to ninth position) that, in practice, become indistinguishable. 

Given the ensemble of ordering protocols performing at 95\% the optimal one, we calculate the fraction of age groups that appear in the first, second, and third vaccinating position. For example, if the ensemble contains only two protocols, say $[9,8,7,6,5,4,3,2,1]$ and $[8,9,7,6,5,4,3,2,1]$, groups 9 and 8 appear once in first and once in second position, and group 7 appears twice in the third position. Fig.~\ref{fig:123Unconstrained} summarizes group prioritization for these three positions and each vaccination rate. Let us examine each vaccination rate in turn. 

For $v=0.05$, as described, epidemic dynamics are dominated by free contagion from infected to susceptible individuals, and the flux from susceptible to recovered due to vaccination is mostly inconsequential: contacts do not play any major role, reduction of casualties is maximal if the prioritized group is the eldest, and reduction in the number of infections (obtained through direct vaccination) does not depend significantly on which group is vaccinated first. As shown in Table~\ref{table:optimalUnconstrained}, the RI of the best protocol is only 1.1\%, while the RI of the strict age order (SAO) protocol achieves 1.0\%, representing a minor difference.

However, prioritizing vaccination for the eldest results in the smallest reduction in infection numbers. This is because this group has the highest mortality rate, making vaccination less effective at lowering infection rates; it is also the group with fewer contacts.
Such minimal influence of the eldest group in the RI is reflected in Fig.~\ref{fig:123Unconstrained}. For $v=0.05$ there is no strategy prioritizing this group included among the best performing RI protocols (RI within the optimal, $1.1$, and 5\% variation, 1.045).

This explains why in Fig.~\ref{fig:123Unconstrained}, lower panel, group 9 is absent from the first position, while the advantage of beginning with any other group yields comparable reductions in RI. 

\begin{widetext}
\begin{center}
\begin{figure}[H]
    \centering
    \includegraphics[width=0.8\textwidth]{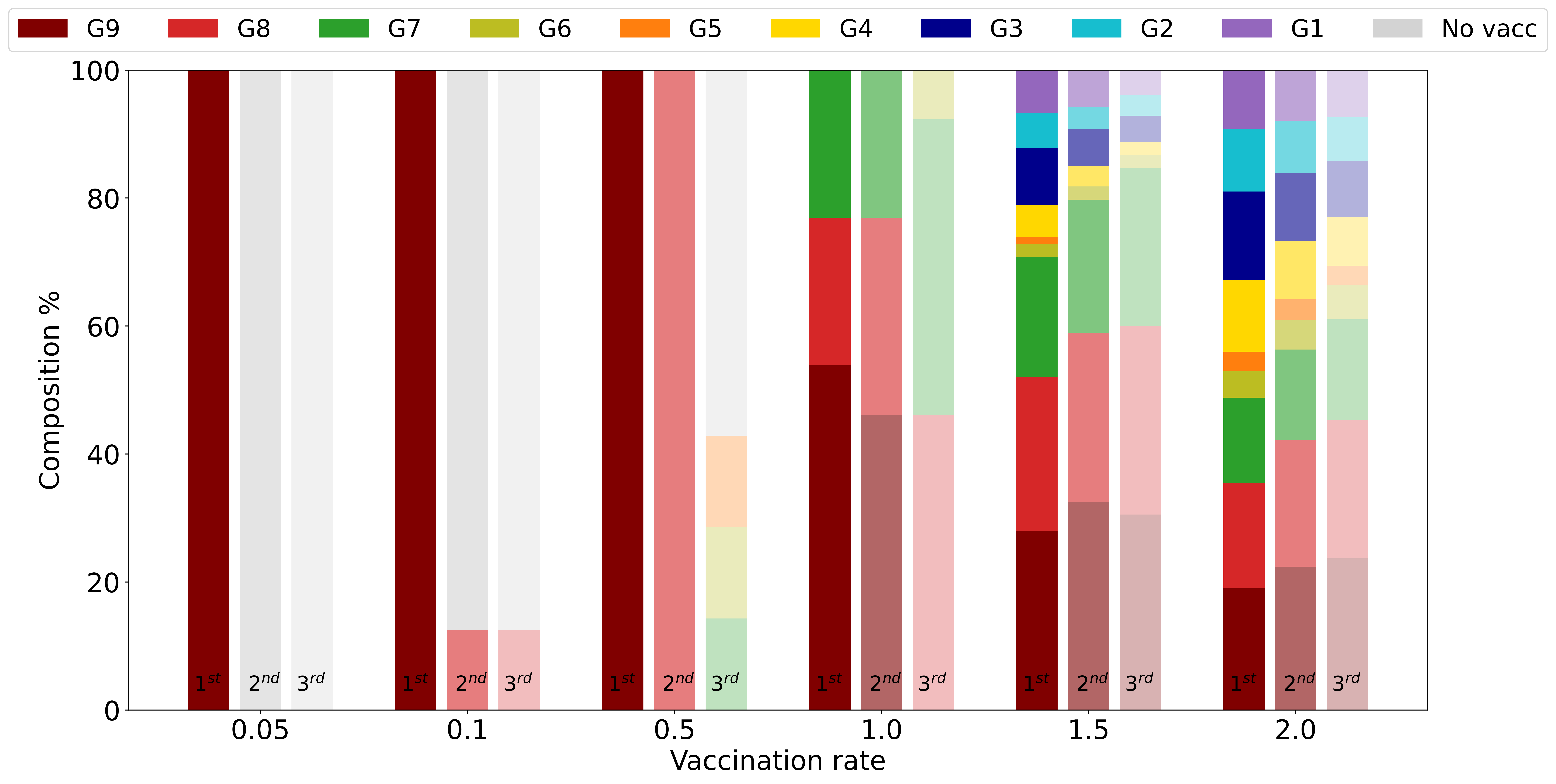}
    \includegraphics[width=0.8\textwidth]{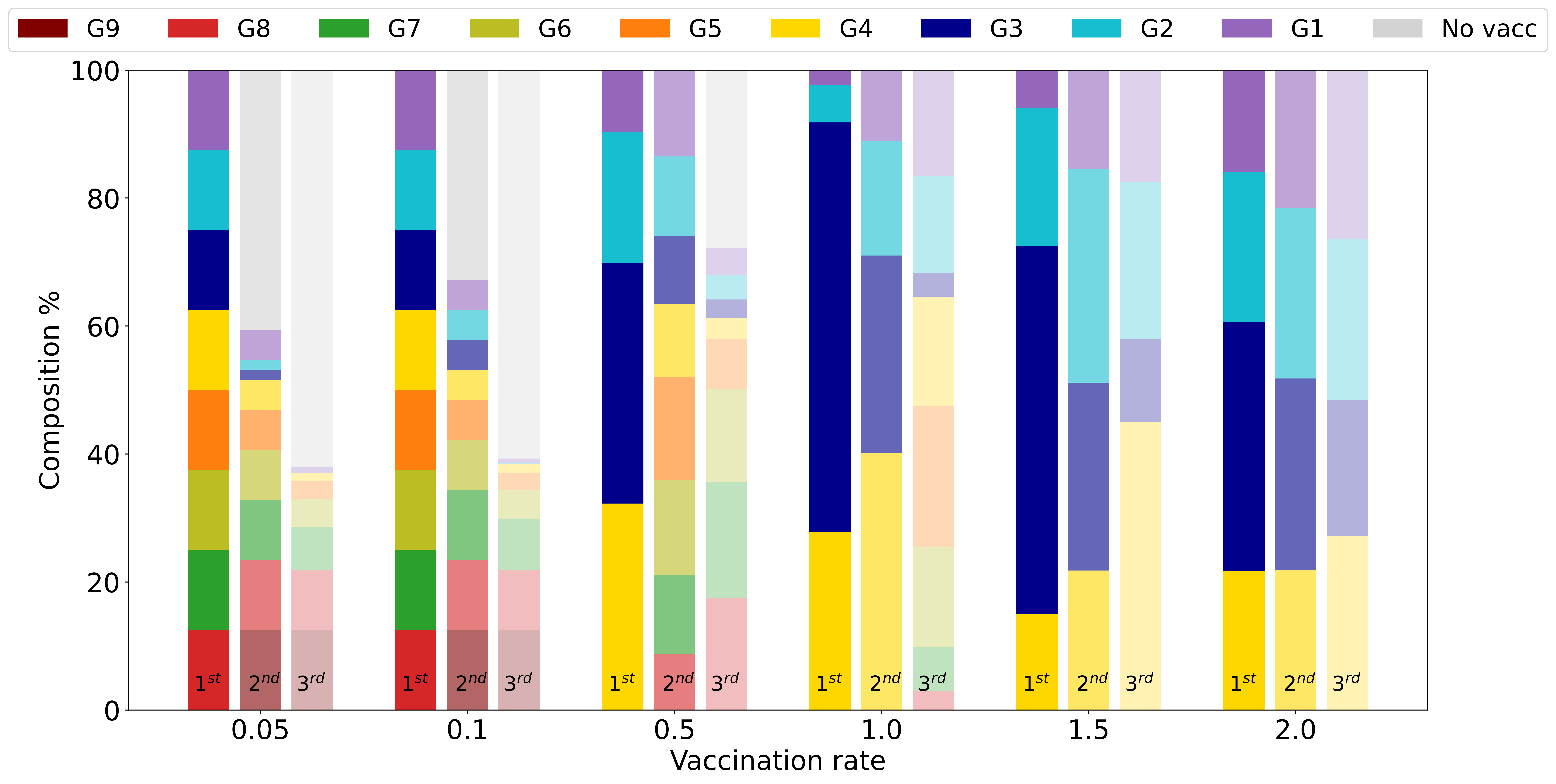}
    \caption{First, second, and third position distribution for each vaccination rate considering top protocols with a level of performance higher than 95\% of the best one in terms of (upper) RD and (lower) RI. The colors indicated are for the first position; second has a lighter shade, and third even lighter.}
    \label{fig:123Unconstrained}
\end{figure}
\end{center}
\end{widetext}

Quantitative changes appear for $v=0.1$, Fig.~\ref{fig:violinUnconstrained}, since the reduction in the fraction of casualties reaches 18.5\% for the best protocol, which is the strict age order protocol $[9,8,7,6,5,4,3,2,1]$. When all protocols causing reductions of 17.58\% or higher are considered, up to three groups can be vaccinated, but a strong preference for protocol $[9,8,0,0,0,0,0,0,0]$ is observed: only strategies vaccinating in this age order are able to perform among the best 5\% in RD.
As for RI, the situation is very similar to the previous vaccination rate. Contacts do not affect the dynamics in any significant way, which continues to be dominated by free propagation plus the direct effect of vaccination: twice the number of vaccines doubles the reduction in the number of infections, with an irrelevant dependence on the vaccination order.  

The situation changes qualitatively for $v=0.5$, where signs of a non-linear dependence between vaccination and observed benefits show up. The difference in performance among protocols becomes larger for RD, while all of them still perform similarly for RI, Fig.~\ref{fig:violinUnconstrained}. For RD, the SAO protocol continues to be the preferred option, now allowing the vaccination of three groups. Protocols with groups 7, 6 and 5 in the third position are found among the best 95\% orderings (that is, causing a reduction of at least 51.87\% in RD). Interestingly, some protocols that begin with the vaccination of the second-oldest group, as $[8,9,7,0,0,0,0,0,0]$ perform now better, regarding RD, than some protocols that begin with the oldest group, as $[9,7,8,0,0,0,0,0,0]$.

Together with the previous rate, $v=1.0$ is the rate where the effect of the order of vaccination has the strongest effect. The difference between the optimal and the worst protocol raises to about 50\% for RD; still, the dispersion of RI values remains small. At the same time, protocols different from SAO offer comparable advantages, and even protocols that begin with vaccination of groups 1, 2 or 3 reach RD values over 60\%. Optimal reductions in the number of infections are achieved when vaccination starts with groups of intermediate ages, followed by children and leaving the eldest groups at the end of the protocol. These orderings, however, perform very poorly regarding RD, so they are not an option for COVID-19 at this vaccination rate. 

When $v$ increases further, the effects of contacts, IFR and vaccination rate intermingle in a complex way, eventually leading to orderings that turn out to be optimal both for RD and RI when vaccination begins with groups 3, 2, and 4. At this point, the spread of RD values diminishes, and vaccination speed overcomes the basal dynamics of contagion, yielding an advantage when more connected groups are vaccinated first. This trend continues for values of the vaccination rate above $v=2.0$, the largest value used here in our simulations. 

\subsection{Optimal ordering under social contact constraints}
Limitations in social contacts, reducing
the average number of contacts an individual has, should impact disease transmission. As explained in section \ref{sec:constrainedcontacts}, we have used two empirical studies carried out under lockdown measures to estimate contact reduction between all age-group pairs. This allows us to quantify the joint effect of vaccination and social contact
restrictions and to compare the result with the baseline case in the previous section. In the case of COVID-19, we observe that the effect the union of the two dissimilar mechanisms may have had in its spread is not only quantitative. According to our model, and as we show in the following, the combination of an optimal vaccination protocol with a sufficient reduction in the number of effective contacts may have caused locally a halt of epidemic propagation. An illustration of this effect through epidemic dynamics of the model can be seen in Figures S1-S4, in the Supplementary Material. 

\subsubsection{Reduction in fatalities and infections}
\begin{figure}[t]
    \centering
    \includegraphics[width=0.4\textwidth]{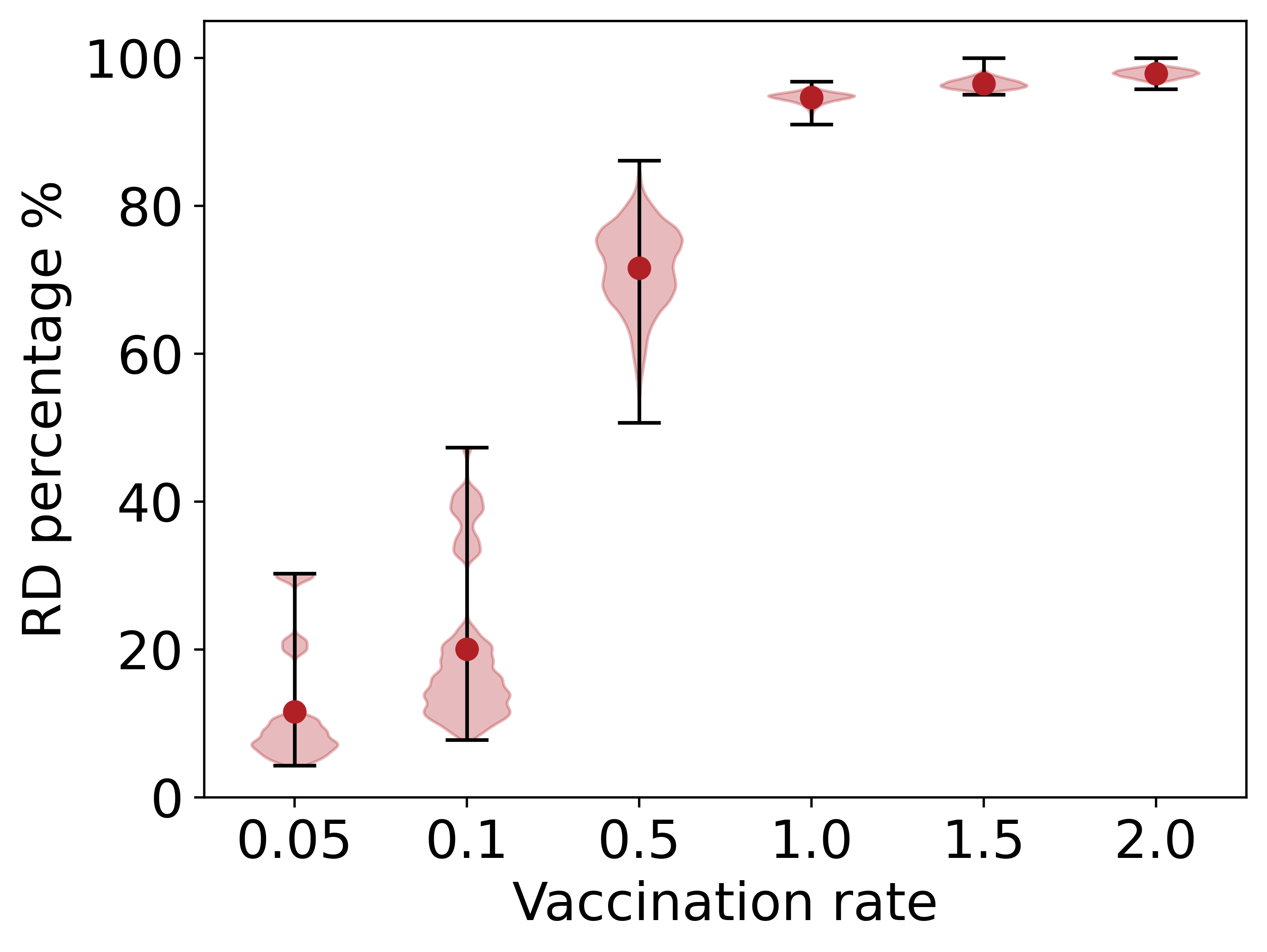}
    \includegraphics[width=0.4\textwidth]{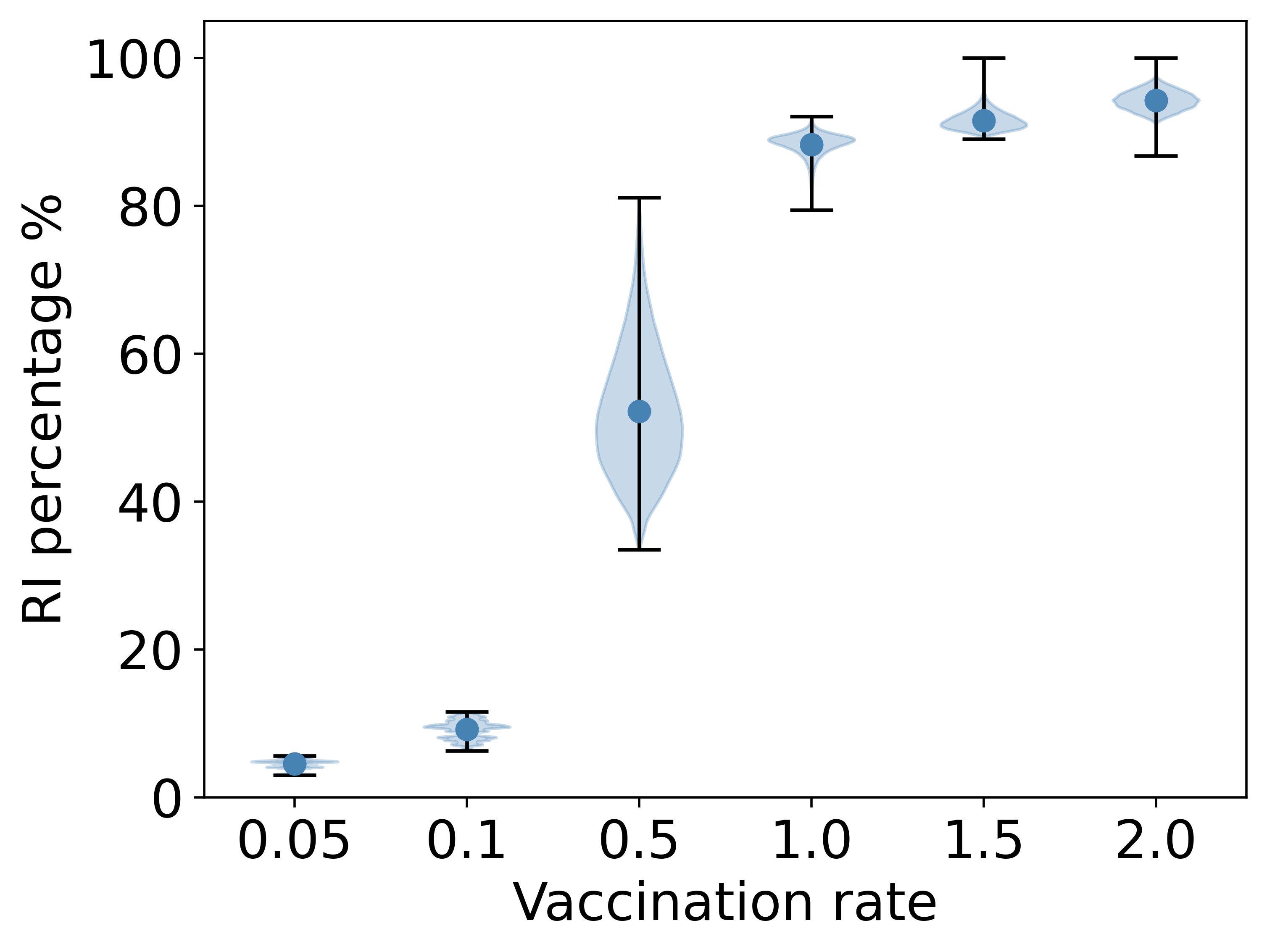}
    \caption{Distribution of the percentage reduction in (top) deaths and (bottom) infections for all possible permutations of the vaccination protocol, for each vaccination rate, in the case with restricted contacts. Dots represent average values, and black bars span from maximum to minimum effect.}
    \label{fig:violinConstrained}
\end{figure}

As in the previous case, we analyze the distribution of RD\% and RI\% considering all possible orderings at each vaccination rate. The results are represented as violin plots in Fig.~\ref{fig:violinConstrained} and should be compared with those in Fig.~\ref{fig:violinUnconstrained}.

Though general considerations are similar to those discussed for the case with unconstrained contacts, two main differences emerge. First, a significant dispersion in RD\% values occurs even at the lowest vaccination rates, highlighting on the one hand a non-linear interaction among adjustable variables (contact limitation and vaccination rate) and, on the other hand, the relevance of properly selecting the optimal vaccination order. With reduced contacts, infection progresses slower, and therefore, there is more ``time to act'', explaining why there can be bigger differences among strategies. Secondly, even at intermediate vaccination rates ($0.5 \le v \le 1.0$) RD\% and RI\% values attained are above the fraction of vaccinated individuals; for $v \ge 1.5$ both infection propagation and mortality are fully suppressed. This indicates that, for $v \ge 1$, epidemic propagation can be halted through a suitable election of vaccination ordering and measures to limit social contact. Interestingly, non-linear effects produce a relatively fast transition from vaccination rates at which there is a direct proportionality between the number of vaccinated individuals and the total reduction in fatalities plus infections, and vaccination rates able to fully inhibit infection propagation. 

\subsubsection{Optimal protocols substantially differ from older-to-younger ordering}

A significant reduction of contacts between individuals at intermediate vaccination rates causes a full inhibition of infection propagation due to non-linear effects between contact reduction and vaccination. The selection of the optimal vaccination order, once the two latter variables are fixed, becomes essential to maximize their synergistic effect. Fig.~\ref{fig:123Constrained} and Table~\ref{table:redMx_bestRD_RI} summarize the results regarding optimal orderings. Remarkably, even at intermediate vaccination rates the optimal ordering substantially differs from the baseline, SAO ordering. At $v=0.5$ vaccination should ideally start with the youngest group, followed by groups 4 and 3, to yield an improvement in RD\% of about 10\% with respect to SAO ordering; under the same protocol, the improvement in RI\% more than doubles. Remarkably, both variables are significantly improved under protocols that coincide in the first four groups to be vaccinated (see Table~\ref{table:redMx_bestRD_RI}). The epidemic is practically fully contained at vaccination rates $v=1.0$ and higher.

Note that the response of the different groups and of the whole population at a given vaccination rate under limited social contacts
is not a simple translation of that obtained at higher vaccination rates under unlimited contacts, as comparison of Figures \ref{fig:123Unconstrained} and \ref{fig:123Constrained} shows (see also Supplementary Figures S1-S4). One reason are the changes in the contact matrix, with different group ages leading the number of contacts under the two different conditions analyzed. In particular, it is only in a situation with limitations in social contacts that the vaccination of the youngest and intermediate age groups yields high advantages in both RD\% and RI\% and is therefore justified.

\begin{table}[t]
\centering
    \caption{Optimal vaccination orderings to minimize the number of deaths and the number of infections under contact restrictions. Table structure as in \ref{table:optimalUnconstrained}. Best performing strategies are compared with the age-ordered baseline protocol.}
    \label{table:redMx_bestRD_RI}
    \begin{tabular}{|c|l|c|r|r|r|r|}
        \hline
        {\bf } & $v$ & Protocol & RD\% & RI\% & SAO RD\% & SAO RI\% \\
        \hline 
        \multirow{6}{0.75em}{\rotatebox[origin=c]{90}{Best RD}}& $0.05$ & [9 8 0 0 0 0 0 0 0] & $30.27$ & $3.01$ & $30.27$ & $3.01$ \\
        & $0.1$ & [9 8 7 0 0 0 0 0 0] & $47.33$ & $6.30$ & $47.33$ & $6.30$ \\ 
        & $0.5$ & [1 4 3 2 9 8 6 7 5] & $86.11$ & $74.67$ & $77.42$ & $33.62$ \\
        & $1.0$ & [1 4 3 2 5 6 8 7 9] & $96.80$ & $92.06$ & $91.71$ & $80.03$ \\ 
        & $1.5$ & [4 5 3 2 8 9 7 6 1] & $99.99$ & $95.72$ & $95.18$ & $89.35$ \\
        & $2.0$ & [3 1 2 6 8 4 9 7 5] & $99.99$ & $98.46$ & $96.37$ & $91.29$ \\ \hline \hline
        \multirow{6}{0.75em}{\rotatebox[origin=c]{90}{Best RI}} & $0.05$ & [4 3 5 6 7 8 9 0 0] & $6.34$ & $5.59$ & $30.27$ & $3.01$ \\ 
        & $0.1$ & [4 3 5 6 7 8 9 0 0] & $12.16$ & $11.57$ & $47.33$ & $6.30$ \\ 
        & $0.5$ & [1 4 3 2 5 6 7 8 9] & $80.70$ & $81.11$ & $77.42$ & $33.62$ \\
        & $1.0$ & [1 3 4 2 5 6 7 8 9] & $96.79$ & $92.07$ & $91.71$ & $80.03$ \\ 
        & $1.5$ & [2 6 5 3 4 8 1 7 9] & $99.99$ & $99.98$ & $95.18$ & $89.35$ \\
        & $2.0$ & [3 1 2 6 8 4 7 9 5] & $99.99$ & $99.99$ & $96.37$ & $91.29$ \\ 
        \hline
    \end{tabular}   
\end{table}

\begin{widetext}
\begin{center}
\begin{figure}[t]
    \centering
    \includegraphics[width=0.8\textwidth]{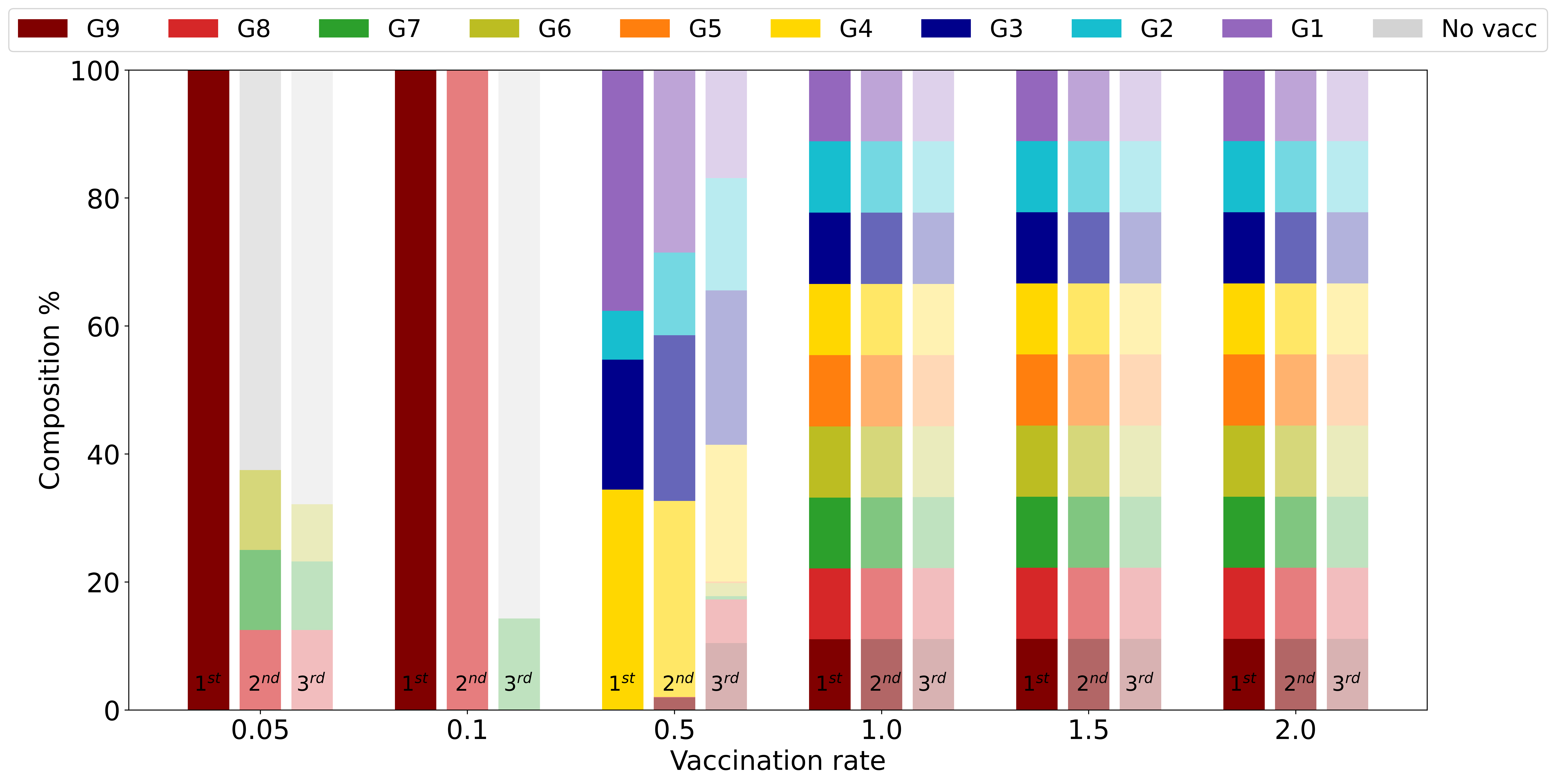}
    \includegraphics[width=0.8\textwidth]{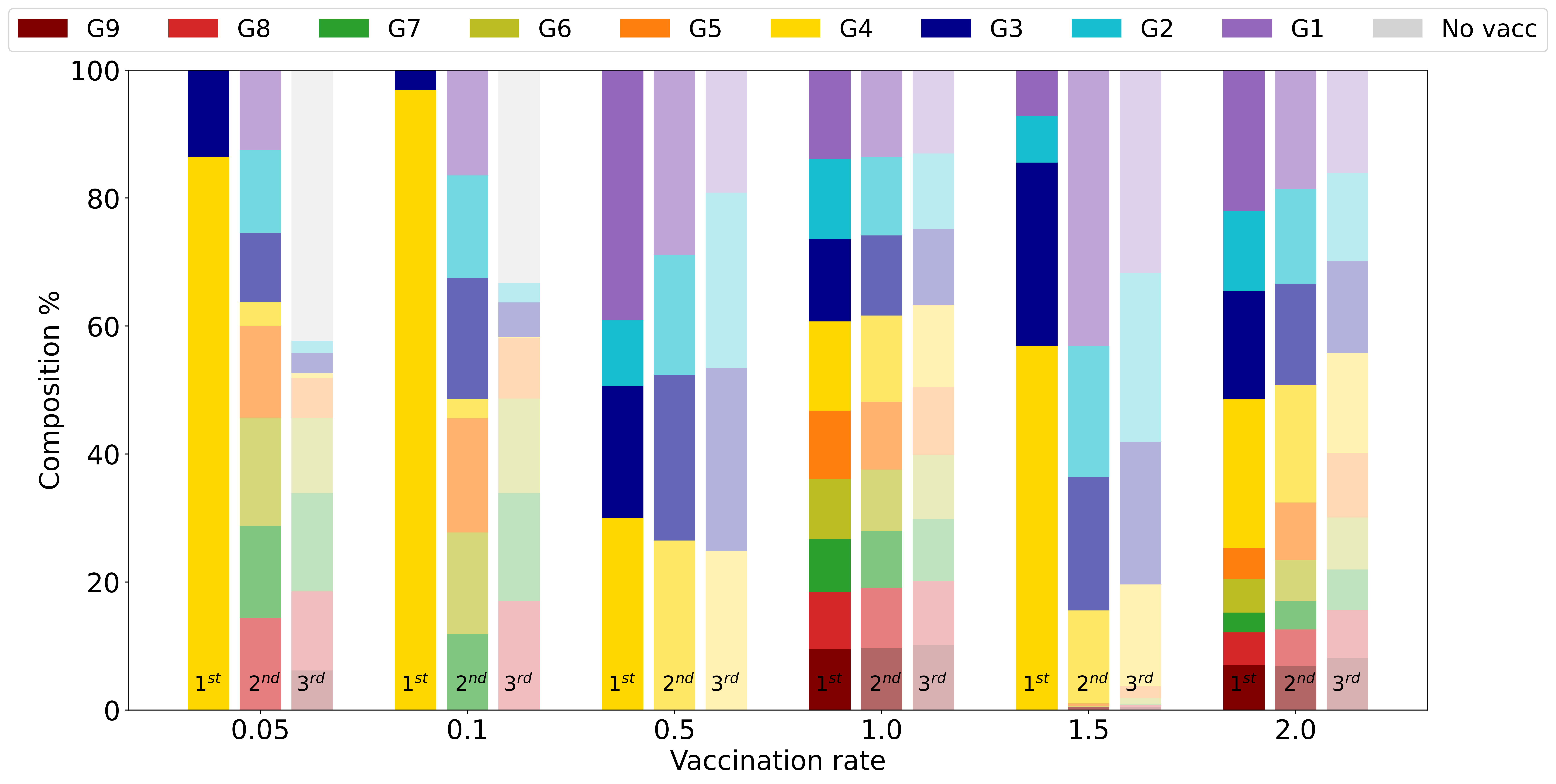}
    \caption{First, second and third position distribution for each vaccination rate considering top protocols with a level of performance higher than 95\% of the best one in terms of (upper) RD\% and (lower) RI\% under restricted contacts. The colors indicated are for the first position; second has a lighter shade, and third even lighter.}
    \label{fig:123Constrained}
\end{figure}
\end{center}
\end{widetext}

\section{Discussion}

In this study, we conducted an exhaustive exploration of all possible vaccine administration protocols across nine age groups, six vaccination rates, and two scenarios informed by field data, with and without social contact reductions. Using the demographic structure of the Spanish population, we simulated epidemic dynamics through a model that includes reinfections, potentially leading to long-term endemic states \cite{rodriguez-maroto:2023}, as observed in COVID-19. Our findings highlight that the effectiveness of vaccination protocols is highly dependent on vaccination rates and the implementation of measures to limit social contacts. 

With unrestricted social contacts, a vaccination rate of 1.0\% of the population per day makes a strict older-to-younger vaccination order the optimal strategy to minimize casualties. However, if the vaccination rate increases to 2\%, prioritizing the 20-29 age group, followed by the 10-19 and 30-39 age groups, becomes more effective. Still, even at the highest explored vaccination rates ($v=2.0$), prioritizing younger groups never yields more than a 4\% reduction in RD without limiting social contacts. This quantitative advantage, while notable, should be interpreted cautiously, given that our model excludes other potentially relevant factors, such as sex-dependent responses to infection, age-dependent vaccination efficacy, values different from 70\% maximum vaccination within each age group, or variations in the value of transmissibility. Nonetheless, the fact that some countries or regions experienced vaccination rates exceeding 2\% per day underscores the potential of alternative protocols for minimizing fatalities, even if the improvements are limited. Other studies have presented results qualitatively consistent with our findings when differential allocation of vaccine doses to age groups is considered. For instance, a study on the Delta variant's spread in Australia demonstrated the benefits of prioritizing socially active groups for vaccination over vulnerable groups, showing greater benefits as vaccination coverage increased \cite{mcbryde:2021}. These results align with findings for other contagious diseases; in the case of influenza, vaccinating schoolchildren and adults aged 30 to 39 offers significant protection by reducing transmission from the most active spreaders to the wider population \cite{medlock:2009}.
However, caution is advised, since in our work we have not considered the existence of asymptomatic individuals for which the infection goes undetected.
Some studies taking this aspect into account have highlighted the risks of high vaccination rates in diseases with a high proportion of asymptomatic individuals, who can unknowingly contribute to the spread of infection \cite{grinfeld:2024}.

The situation changes when social contacts are restricted. Our results demonstrate a complex interaction between social contact patterns and vaccination rates, which directly influences the optimal vaccine rollout strategy. Preference for an older-to-younger vaccination order shifts under a lockdown scenario: at a 1.0\% vaccination rate, prioritizing the youngest age group (0-9 years old), followed by the 30-39 and 20-29 groups, with the oldest group (over 80) last, is optimal. If the vaccination rate increases to 2\%, the recommended order changes to 20-29, followed by 0-9 and 10-19, leaving the 40-49 group in the last position. 

Increasing vaccination rates and limiting social contacts are distinct and non-equivalent strategies for controlling the spread of contagious diseases. The effectiveness of each strategy varies both quantitatively and qualitatively across different age groups. This finding aligns with a mathematical model analyzing the reopening of New Zealand's borders, which indicates that even with a highly effective vaccine, additional public health measures are necessary to manage COVID-19 risks \cite{nguyen:2021}.
Similarly, work using agent-based models has shown a positive synergistic effect between mild and social restrictions and intermediate vaccination rates in controlling epidemics \cite{spiliotis:2022}.
Another study, using COVID-19 dynamics with demographics similar to the general U.S. population, found that if no measures to limit social contacts are implemented, prioritizing older age groups is advisable when vaccine effectiveness or supply is low. However, when both are high, prioritizing younger age groups is more effective due to their role in driving transmission \cite{matrajt:2021}.

For diseases with infection fatality rates varying by age group, incorporating both vaccination strategies and contact reduction measures simultaneously could be beneficial, although direct evidence is still limited. A modeling study comparing the propagation of COVID-19 and influenza using demographic data from Brazil, Uganda, and Germany found that early and widespread vaccination, combined with measures to reduce social contacts, was crucial for reducing deaths. This approach proved to be more important than the specific vaccination strategy employed \cite{schulenburg:2022}.

While our study primarily examines the Spanish population, its analysis readily extends to diverse countries or regions. Critical vaccination rates, triggering a shift in age-group prioritization, depend on demographic distribution and contact habits. Prior findings \cite{rodriguez-maroto:2023} suggest that countries or regions with younger populations or increased contact with average-age groups may benefit from vaccinating younger groups at lower rates.
It is crucial to emphasize that any quantitative advantage hinges on demographics, vaccination rates, vaccine effectiveness, disease specific infection fatality rate, and additional epidemiological variables.
Consequently, no absolute recommendation can be made for optimal vaccination protocols, and each case needs a tailored analysis considering its unique characteristics, as it has been already pointed out \cite{schulenburg:2022,gonzalez-parra:2024}.

Under social contact restrictions, the benefit of prioritizing vaccination for groups with higher social contacts becomes more pronounced, highlighting the importance of considering protocols that focus on younger or more socially active groups. It's crucial to recognize that the value of these alternative vaccination orderings extends beyond just the reduction in deaths and infections. During severe lockdowns, vaccination at intermediate rates—achieved globally during the COVID-19 pandemic—can significantly slow the spread of the virus. This not only reduces the total number of infections but also flattens the curve of active cases, thereby easing the burden on healthcare systems.
While a primary goal of COVID-19 vaccination is to minimize fatalities \cite{european:2021,stiko:2023},
reducing infections is equally important, particularly in preventing the emergence of new variants \cite{zhou:2021,flower:2021,haque:2022}. The optimal vaccination strategy differs depending on the desired outcome:
for instance, different strategies minimize fatalities or infections. Although an approach minimizing infections may not produce the most desirable results in terms of public health for some diseases where some groups have high fatality rates under unrestricted social contacts,
it is effective for illnesses like the flu,
where children are often vaccinated first \cite{medlock:2009,rodriguez-maroto:2023,website:NSWdata,govaert:1994}.
For any contagious disease, making informed decisions about vaccination prioritization is crucial, as it impacts both mortality and infection rates, as demonstrated by the models used in this study.

\section*{Acknowledgments}

The authors are indebted to A. Vespignani for support with the use of their data.
This research was supported through grants PID2020-113284GB-C21 (I.A-D., S.M.), PID2019-109320GB-100, (BADS, G.R-M., S.A.), and PID2022-142185NB-C21 (PGE, S.A.), funded by
MICIU/AEI/10.13039/501100011033 and by ``ERDF/EU'' (PGE). MICIU/AEI/10.13039/501100011033 has also funded the ``Severo Ochoa'' Centers of Excellence to CNB, CEX2023-001386-S, and the special grant PIE 2020-20E079 (CNB) entitled ``Development of protection strategies against SARS-CoV-2''.


\appendix

\section[\appendixname~\thesection]{Calculation of contact matrices}
\label{appendix:M}

As introduced in the main text, our main objective was to study the effects of vaccination by comparing a \textit{``normal"} situation, where contacts are not limited, with a \textit{``confinement"} scenario in which contacts restrictions exist. Data on contact reduction under lockdown for Spain were not available, however, so we resorted to two independent studies for the Netherlands~\cite{backer:2021} and England~\cite{gimma:2022} to estimate a realistic matrix under contact reduction. We used both studies as a guideline in our particular case in Spain, as they compared the reduced matrix to a baseline, pre-pandemic situation. However, data can not be directly applied to our case. We need to process their original contact matrices in order to meet the age grouping used in our study. Reduced contact matrices can be found in the file SuppFile\_ConstrainedMatrix.xlsx, available at \cite{Atienza2024CodeZenodo}.

\subsection{Dutch population study (\textit{NLD\_Backer2021)}
}
We use two of their contact matrices corresponding to data collected from February 2016 to October 2017 (\textit{Baseline}) and April 2020 (\textit{Lockdown}), respectively. Both matrices consider the same age-grouping, so they are comparable with each other. However, one of our original 10-year groups is missing, G1 [0,10) years, which is divided into two smaller subgroups, [0,4] and [5,9]. We need to group all the information of their contacts, either among themselves or with the rest of the groups, in order to compare it later with the original matrix without contact restrictions \cite{mistry:2021}. Considering the structure of the matrix, $M_{1j}$ correspond to the contacts that G1 individuals have towards the other groups. We calculate these values considering the smaller subgroups of the study $\forall j\neq1$ as follows:

\begin{equation}
    M_{1j} = \frac{(S_{(0-4)j}*P_{0-4}) + (S_{(5-9)j}*P_{5-9}) }{P_{0-4}+P_{5-9}}
    \label{eq:outcoming_contacts}
\end{equation}
where $S_{(0-4)j}$ and $S_{(5-9)j}$ are the contacts of the original subgroups with the $j$ group in the matrix of the study and $P_{0-4}$, $P_{5-9}$ the number of participants in the survey for each subgroup. 

In the particular case, $j=1$ we re-calculate the contacts within the group:

\begin{widetext}
\begin{equation}
    M_{11} = \frac{((S_{(0-4)(0-4)}+S_{(0-4)(5-9)})*P_{0-4}) + ((S_{(5-9)(0-4)}+S_{(5-9)(5-9)})*P_{5-9}) }{P_{0-4}+P_{5-9}}
    \label{eq:internal_contacts}
\end{equation}
\end{widetext}

which accounts for all contacts that individuals from groups [0,4] and [5,9] have with individuals from the other two groups. This is because both subgroups are now part of the same \textit{``major"} group [0,10).

The elements $M_{i1}$ therefore represent the contacts that other groups have with the individuals of the new group [0,10). In this case, we just have to add the contacts with the original two subgroups: 

\begin{equation}
    M_{i1} = S_{i(0-4)}+S_{i(5-9)}
    \label{eq:incoming_contacts}
\end{equation}

We apply this transformation for both matrices in the study obtaining the corresponding $9\times9$ matrices in the pre-pandemic (\textit{Baseline}, $B$) and lockdown (\textit{April 20}, $A$) situations; matrices available in the file SuppFile\_ConstrainedMatrix.xlsx at \cite{Atienza2024CodeZenodo}. The green area in both matrices highlights the region that remains constant with respect to the original matrix. Once their matrices are updated to consider our 10-year age groups we calculate the reduction factor matrix, $R$, as the ratio between lockdown and pre-pandemic contacts:

\begin{equation}
    R_{ij} = \frac{A_{ij}}{B_{ij}}
    \label{eq:reduction_factor}
\end{equation}

\subsection{English population study (\textit{GBR\_Gimma2022})}

In this study, the transformation of the data is more complex because the age groups defined are different when comparing the baseline and lockdown matrices.

\begin{itemize}
    \item \textit{Baseline}: The POLYMOD \cite{mossong:2008} matrix is $15\times15$, as we have 5-years intervals for individuals aged $<70$ and a final group with contact information for individuals over $70$. In this case the transformation is simple: We apply the process described in equations \ref{eq:outcoming_contacts}, \ref{eq:internal_contacts} and \ref{eq:incoming_contacts}, considering 5-year age groups two by two, to compute the outcoming, internal and incoming contacts for each original 10-year group. 

    \item \textit{Lockdown}: The distribution of age groups in this $9\times9$ matrix is different compared to the baseline POLYMOD matrix. Four groups are defined for indivudals aged less than $30$: [0,4], [5,11], [12,17] and [18,29]. The problem here is that we can not combine their contact information into any of the original groups [0,9], [10,19] or [20,29] as they are not well-defined by intervals. To solve this issue we have to define some intermediate age groups in order to re-distribute contacts: The [5,11] group is divided into [5,9] and [10,11] and the [18,29] into [18,19] and [20-29] (see \textit{Lockdown1\_inter} matrix in SuppFile\_ConstrainedMatrix.xlsx, available at \cite{Atienza2024CodeZenodo}). Since we lack specific data on the number of individuals of these intermediates, we assume that all ages were equally represented within the groups in the survey. Therefore, their contact contributions are divided proportionally. Considering [5,11] as an example, the intermediate group [5,9] takes information of 5 years within the original group, so it takes $5/7$ of their contacts. Analogous to that, the [10-11] subgroup takes the remaining $2/7$. Following this process, the two intermediates, [18,19] and [20,29], receive $2/12$ and $10/12$ of the [18,29] contacts, respectively. At the end, we recover the original 10-year age groups by adding the corresponding contacts information as previously indicated: G1 [0-10) - [0,4] and [5,9]; G2 [10-20) - [10,11], [12-17] and [18,19] and G3 [20,30) (directly obtained from the second split).    
\end{itemize}

In both matrices, the oldest age group comprises individuals aged 70 years and older. Thus, our two original older age groups, G8 [70,80) and G9 80+, are within it. Since we lack further information on the group over 80 years of age, we assume that the reduction in contacts is the same. Again, we apply the equation \ref{eq:reduction_factor} to obtain the reduction factor matrix. 

\subsection{Spanish matrix reduction}
Finally, we average the two reduction factors ($R_{ij}$ elements) obtained from both studies to multiply the original $M_{ij}$ elements in the pre-pandemic situation in Spain (\textit{Unconstrained SIYRD\_9G}, in SuppFile\_ConstrainedMatrix.xlsx, available at \cite{Atienza2024CodeZenodo}). Thus, we obtain the reduced contact matrix under lockdown in Spain as a case-example of contacts under restrictions (\textit{Constrained SIYRD\_9G} in SuppFile\_ConstrainedMatrix.xlsx). As expected, there is a significant average reduction of $66.9\%$ in the number of daily contacts in Spain. As Table \ref{tab:app3_contactsReduction} illustrates, there is a certain degree of variability between age groups.

\begin{table}[h!]
    \centering
    \begin{tabular}{|c|c|c|c|c|c|c|c|c|c|}
    \hline
    Group & G1 & G2 & G3 & G4 & G5 & G6 & G7 & G8 & G9  \\
    \hline
    \% red. & $59.1$ & $66.6$ & $73.0$ & $66.8$ & $66.0$ & $ 63.7$ & $68.0$ & $68.8$ & $70.0$ \\
    \hline
    \end{tabular}
    \caption[Contacts reduction in Spain]{\textbf{Contacts reduction in Spain.} Percentage reductions are shown compared to a baseline situation without restrictions.}
    \label{tab:app3_contactsReduction}
\end{table}

\end{document}


\vspace*{0.2in}

{\Large
\noindent
{\it Supplementary information to} 

\begin{center}
\noindent
{Optimal COVID-19 Vaccine Prioritization by Age Depends Critically on Inter-group Contacts and Vaccination Rates}

\noindent
{\large Iker Atienza-Diez, Gabriel Rodriguez-Maroto, Sa\'ul Ares and Susanna Manrubia}
\end{center}

}

\section{Dynamics in various scenarios}

Figures S1-S4 show the dynamics of a set of simulations for four different (suboptimal, in general) vaccination protocols implemented with and without contact reduction, and for the six values of the vaccination rate used in this study. Each panel shows the collective dynamics of the five classes in the model, where the curve displays the sum of individuals of all ages in that class. In all cases, the values of percent reduction in deaths and infected individuals (RD, RI) one year after the epidemic started are shown in the inset, using as a baseline the case without vaccination and no restrictions in social contacts. They illustrate how, under our model assumptions and parameter values, herd immunity cannot be attained if social contacts are not reduced, even for high vaccination rates. Under lockdown measures, in contrast, epidemic propagation can be halted even with suboptimal vaccine administration protocols, at intermediate vaccination rates.     

\clearpage
\newpage

\renewcommand{\thefigure}{S\arabic{figure}}

\begin{figure}
\begin{center}
\includegraphics[width=0.92\textwidth]{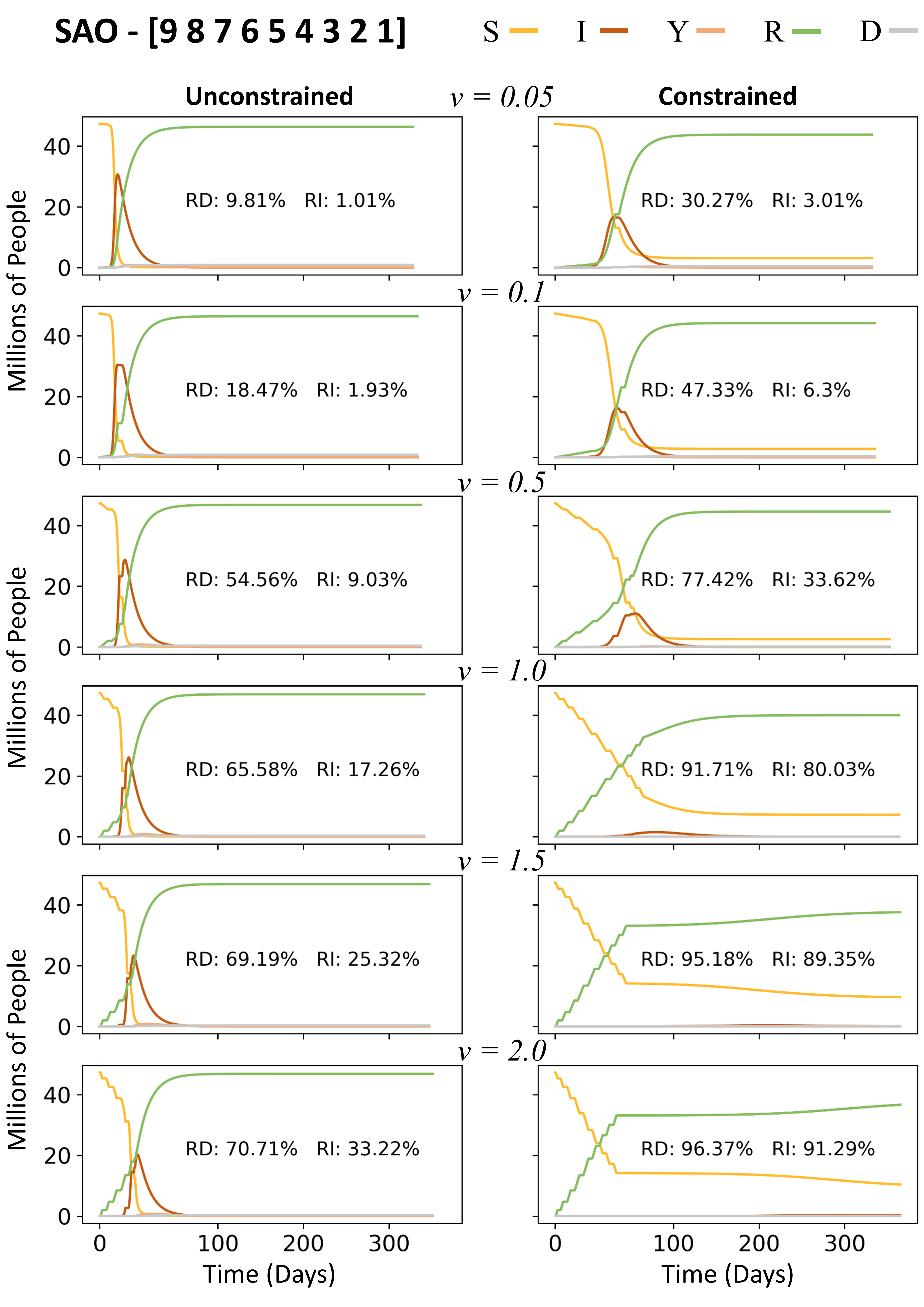}
\caption{Dynamics under the SAO protocol, where vaccination progresses in strict older-to-younger order. The increase in RD is monotonic under unconstrained social contacts, but there is a transition to herd immunity at a vaccination rate $1.0 \le v \le 1.5$ if lockdown measures are enforced. 
}
\label{fig:S1_SAO}
\end{center}
\end{figure}   

\begin{figure}[t]
\begin{center}
\includegraphics[width=0.92\textwidth]{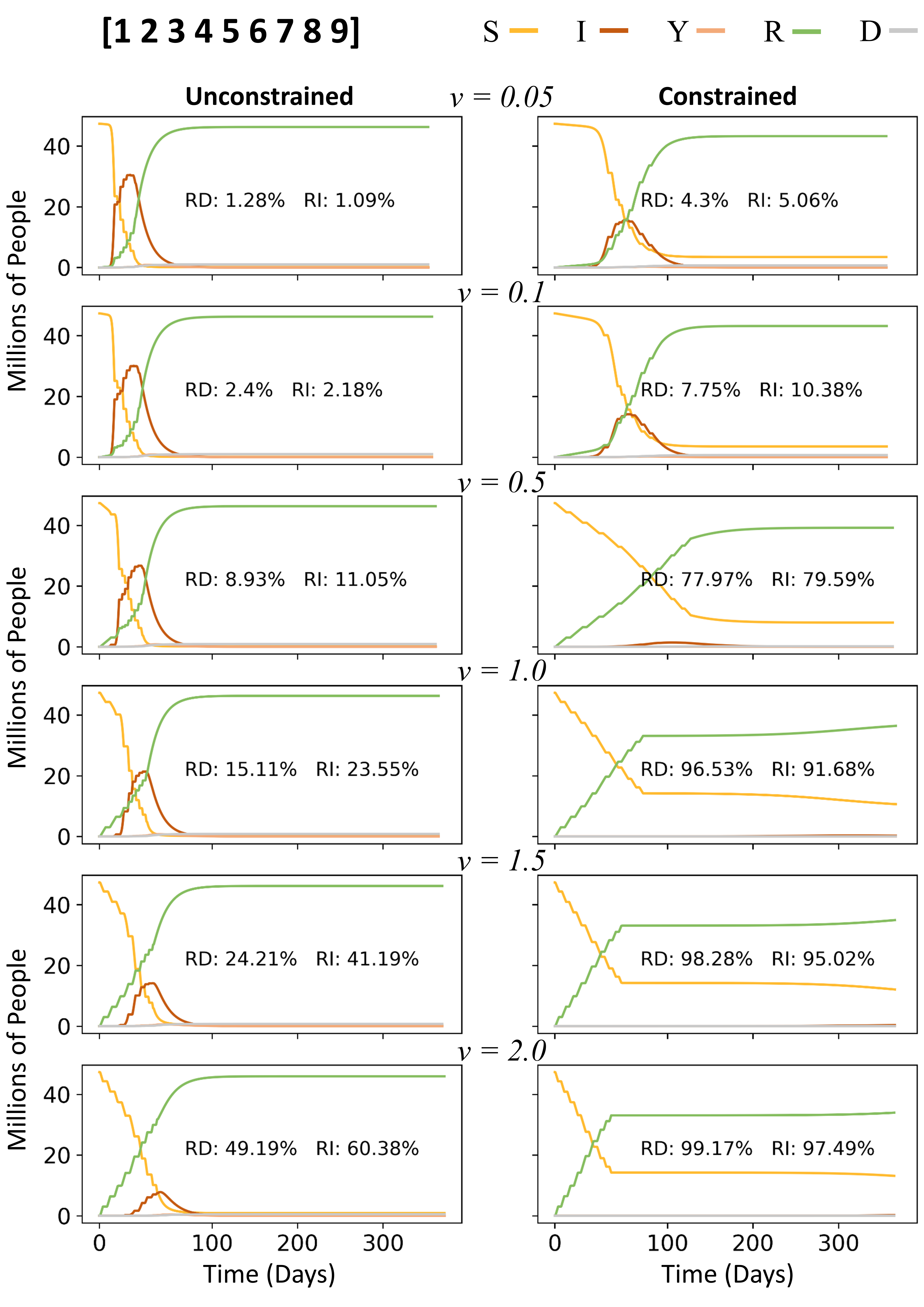}
\caption{Dynamics under a protocol where vaccination progresses in strict younger-to-older order. If no contact restriction is applied, this protocol performs worse than the SAO protocol. However, under lockdown measures this protocol is superior even at moderate vaccination rates, $v \ge 0.5$, and the transition to herd immunity occurs earlier than under the SAO protocol. 
}
\label{fig:S1_invSAO}
\end{center}
\end{figure}   

\begin{figure}[t]
\begin{center}
\includegraphics[width=0.92\textwidth]{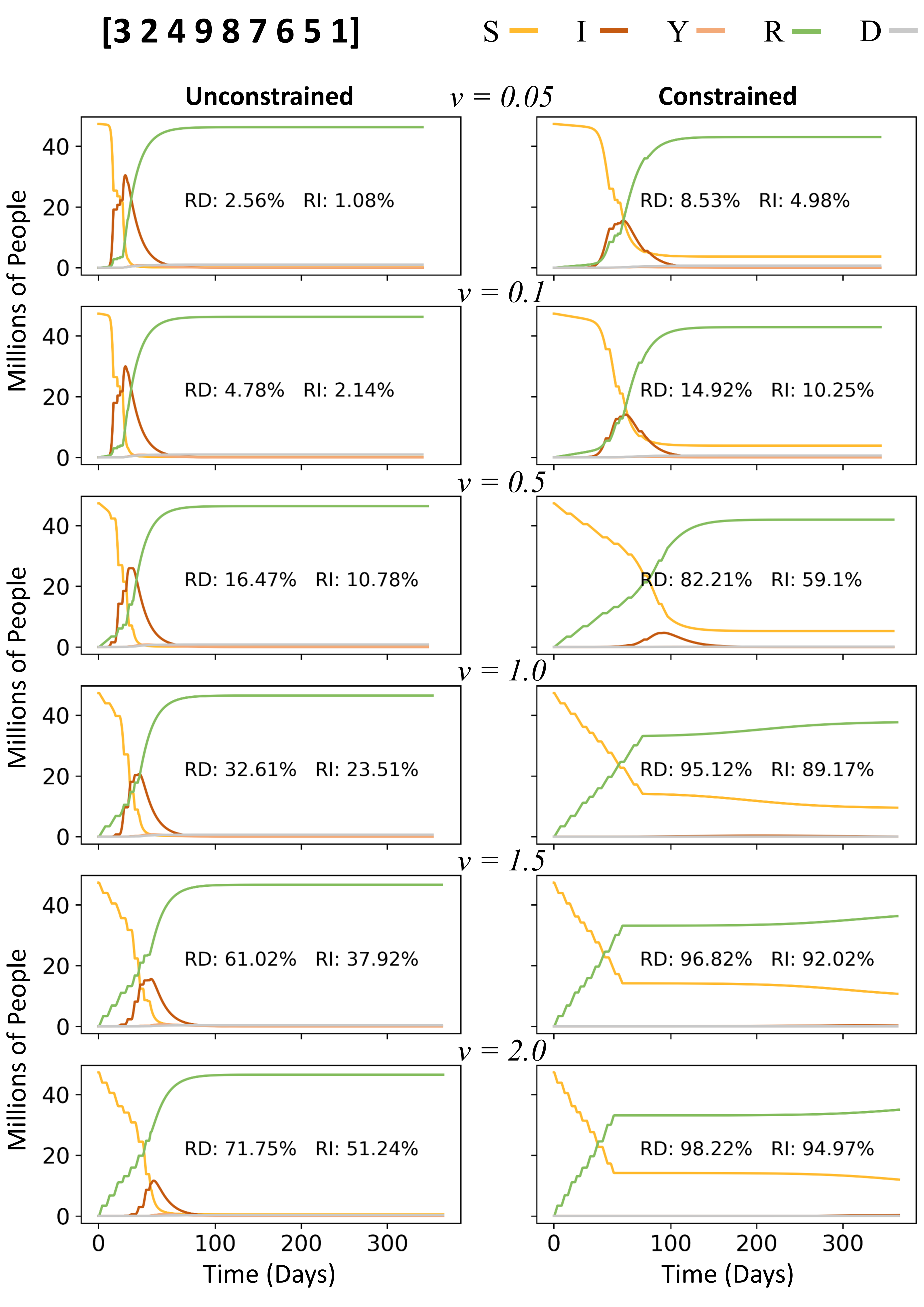}
\caption{Dynamics under the protocol that minimizes the number of deaths under no contact restriction ($v=2.0$, see Table I in the Main Text). At sufficiently high vaccination rates, this protocol performs only slightly better than SAO. 
}
\label{fig:S1_topRD}
\end{center}
\end{figure}   

\begin{figure}[t]
\begin{center}
\includegraphics[width=0.92\textwidth]{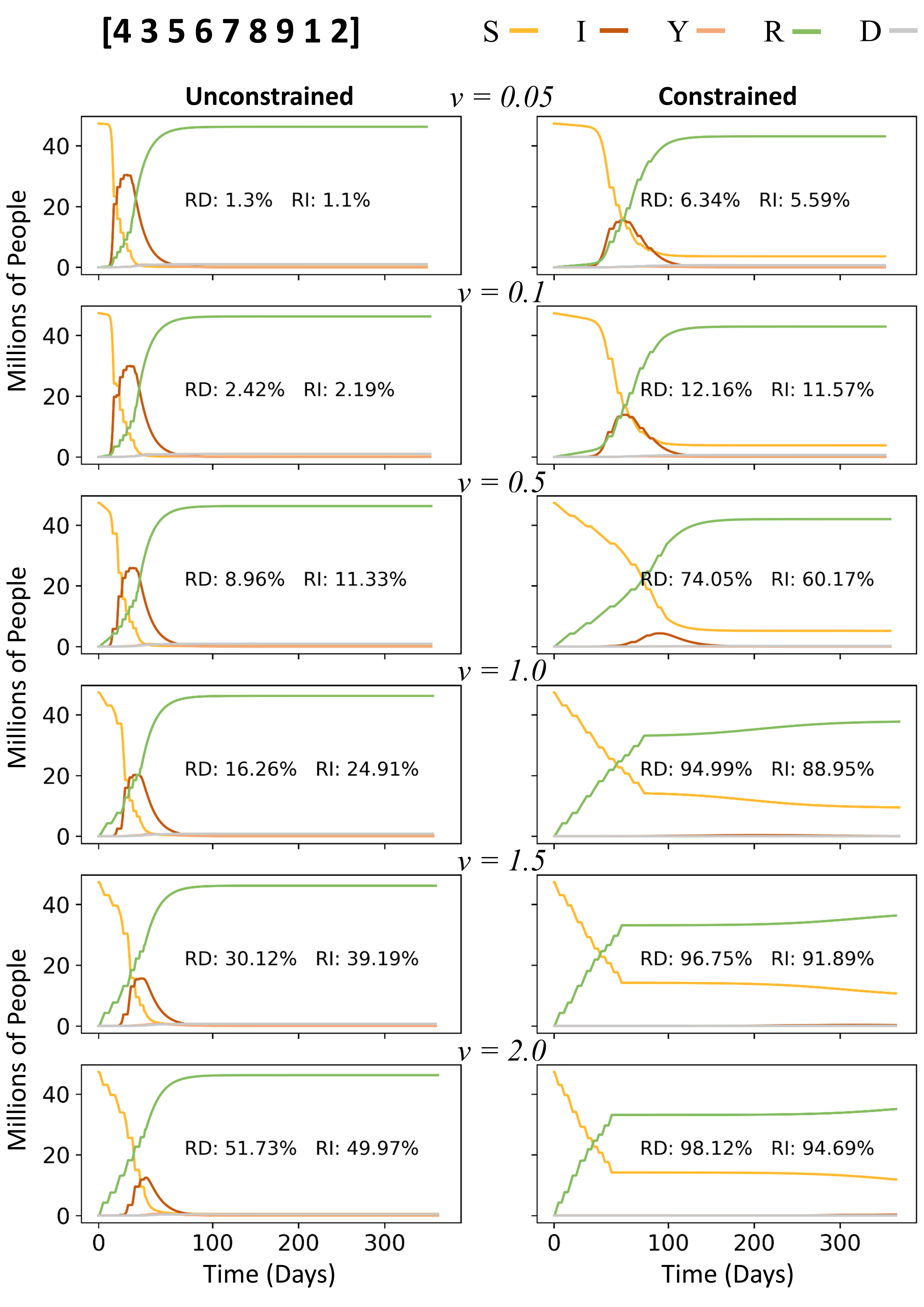}
\caption{Dynamics under the protocol that minimizes the number of infections under contact restriction ($v=0.05$ and $v=0.1$, see Table II in the Main Text.) In general, this protocol performs better than SAO considering RI but not RD. However, under lockdown measures, this protocol is slightly superior with intermediate to high vaccination rates, starting at $v=1.0$. Note that, under the dynamics generating the values in Table II, the last two groups are not vaccinated because 70\% or more of the individuals in groups 1 and 2 are infected before their vaccination turn arrives; however, the dynamics differ in some of the cases shown in this figure, so the complete protocol is in general applied. 
}
\label{fig:S1_topRI}
\end{center}
\end{figure}   
